\documentclass[11pt,preprint,prd,aps,superscriptaddress,nofootinbib,showpacs,a4paper]{revtex4}
\usepackage{color}
\usepackage{graphicx}
\usepackage{epsfig}
\usepackage{amsmath}
\usepackage{amssymb}
\usepackage[normalem]{ulem} 
\newcommand{\be}{\begin{equation}}
\newcommand{\ee}{\end{equation}}
\newcommand{\bqa}{\begin{eqnarray}}
\newcommand{\eqa}{\end{eqnarray}}

\date{\today}
\begin{document} 

\date{\today }

\title{
Amplitude analysis of  \mbox{\boldmath $ B^0 \to K^0_S K^+ K^-$} decays in a quasi-two-body QCD factorization approach
}

\affiliation{Sorbonne Universit\'es, Universit\'e Pierre et Marie Curie,\\
Universit\'e de Paris et IN2P3-CNRS, UMR 7585, \\
 Laboratoire de Physique Nucl\'eaire et de Hautes \'Energies,  4 place Jussieu, 75252 Paris, France}

\affiliation{Division of Theoretical Physics, The Henryk Niewodnicza\'nski Institute of Nuclear Physics,\\
                  Polish Academy of Sciences, 31-342 Krak\'ow, Poland\\}

\author{\bf {J.-P.~Dedonder$^{1*}$}}

\author{\bf {R.~Kami\'nski$^2$}}
                  
\author{\bf {L.~Le\'sniak$^2$}}

\author{\bf {B.~Loiseau$^1$}}

\author{\bf {P.~\.{Z}enczykowski$^2$}}

\thanks{Retired Professor}

\begin{abstract}
\noindent 

The $B^0 \to K^0_S K^+ K^- $ decay amplitude is  derived within a quasi-two-body  QCD factorization framework in terms of kaon form factors and  $B^0$ to two-kaon-transition functions.
The final state kaon-kaon interactions in the $S$, $P$, and $D$ waves are taken into account.
The unitarity constraints are satisfied for the two kaons in scalar states. 
It is shown that with few terms of the full decay amplitude one may reach a fair agreement with the total branching fraction and Dalitz-plot projections published in 2010 by the Belle Collaboration and in 2012 by the \textit{BABAR} Collaboration.
With~13 free parameters, our model fits the corresponding~422 data with a $\chi^2$ of~583.6 which leads to a $\chi^2$ per degree of freedom equal to 1.43.
The dominant branching fraction arises  from the $f_0(K^+K^-) K^0_S$ mode with~83.0~\% of the total branching.
The next important mode is dominated by $\phi K^0_S$ plus small $\omega K^0_S$ and $\rho^0 K^0_S$ modes with~18.3~\% of the total. 
Then follows the $a_0^\pm K^\mp$ mode  with~6.2~\%.
Adding the other smaller modes, the total percentage sum is~107.7~\% which indicates a small interference contribution.
In most regions of the Dalitz plot, our model gives rather small $CP$ asymmetry, but in some parts its values can be large and positive or negative.
Its predicted total value is equal to  $-0.11$~\%.
The calculated time dependent  {\it CP}-asymmetry parameters agree, within errors, with those obtained by the \textit{BABAR} analysis.
Our model amplitude can be the basis for a parametrization in experimental Dalitz plot analyses of LHCb and Belle II Collaborations.

\end{abstract}

\maketitle

\section{Introduction}
\label{Introduction}

The charmless hadronic time dependent $ B^0 \to K^0_S K^+ K^- $ decays have been studied a decade ago by the Belle~\cite{PRD82_073011} and \textit{BABAR}~\cite{PRD85_112010} Collaborations with the aim of extracting {\it CP} violation parameters.
These decays, currently analyzed by the LHCb Collaboration~\cite{Grammatico22},  were used, together with other charmless three-body decays of $B$ mesons, to extract, through Dalitz-plot amplitude analyses, the Cabibbo-Kobayashi-Maskawa (CKM) phase $\gamma$~\cite {Bertholet2019}.
In the experimental analyses the final state meson interactions are  often described by  relativistic Breit-Wigner functions (isobar model) which do not satisfy the unitarity condition\footnote{However, the $S$-wave $f_0(980)$-resonance contribution is fitted though the K-matrix formalism where the two-body unitarity is preserved.}.
The scalar-isovector $a_0$ resonances, present in the $K^0 K^\pm$ final states, are not  introduced in the Belle and \textit{BABAR} analyses. 
This is also the case for  the $\omega$  (mainly $K^+ K^-$ channel) and $\rho$ (mainly $K^0 K^\pm$ channel) resonances.
Belle II Collaboration~\cite{PRD108_072012} has recently measured the variation in time of the rate asymmetries in $ B^0 \to \phi K^0_S$  decays. 
This process, part of the  $B^0 \to K^0_S K^+ K^- $, could reveal some new physics in the $b \to q \bar qs$ transitions.
In these charmless three-body decays, the contribution of diagrams with virtual particle loops is important and consequently their study could exhibit some physics beyond the Standard Model.

 In the method, used by Ref.~\cite {Bertholet2019}, for extracting $\gamma$ from $B \to K \pi \pi$ and $B \to KK\bar K$ reactions, the amplitudes are written as combinations of momentum dependent tree and penguin diagrams with some of them related via the assumed SU(3) flavor symmetry.
There, the model amplitudes, obtained in the different \textit{BABAR} analyses for every studied decay, are taken as experimental inputs.
Among the six possible solutions found for $\gamma$ in Ref.~\cite {Bertholet2019}, one is compatible with the world-average value~\cite{PDG2022} of $\left (65.9^{+3.3}_{-3.5} \right)^\circ$.
The effect of SU(3) symmetry breaking averaged over the  Dalitz plot is calculated to be small.

In Ref.~\cite{ChengPRD76_094006}  charmless three-body decays of $B$ mesons have been thoroughly studied within a quasi-two-body model based on factorization approach.
There, the description of the non-resonant (NR) background, consisting of a point-like weak transition and pole diagrams, is achieved using heavy-meson chiral-perturbation theory.
The momentum dependence of the corresponding amplitudes is assumed to be in the exponential form
to  insure that the predicted decay rates,  in general unexpectedly large, agree reasonably well  with experimental results. 
The final state resonance signals are described in terms of typical relativistic Breit-Wigner expressions.
For the  $B^0 \to K^0_S K^+ K^- $ decay, the branching ratios and the $K^+ K^-$ mass spectra are compared with the available \textit{BABAR} analysis in their Table~III and Figs.~2~(a) and(b), respectively.
The quantum chromodynamic (QCD)  factorized expression for the $B^0 \to K^0_S K^+ K^- $ decay amplitude given by their Eq.~(A4) will be the starting point of our work.

 Taking into account of the Belle~\cite{PRD82_073011, PRD69_012001} and \textit{BABAR}~\cite{PRD85_112010} data, the first two authors of Ref.~\cite{ChengPRD76_094006} have revisited their 2007 model in Ref.~\cite{PRD88_114014} to compare their results with experimental branching fractions and direct {\it CP}-violation in charmless three-body decays of $B$ mesons.
 However, their $B^0 \to K^0_S K^+ K^- $ branching ratio compared to that of \textit{BABAR} is too small.
 These Belle~\cite{PRD69_012001}  and \textit{BABAR} branching values have been recently confirmed by the updated branching fraction measurements of the LHCb Collaboration~\cite{JHEP11(2017)027}.
 
 Let us describe succinctly  some recent studies related to charmless three-body $B$ decays.
A substantial extension of the approach of Refs.~\cite{ChengPRD76_094006} and~\cite{PRD88_114014} has been analyzed in Ref.~\cite{PRD94_094015}.
A perturbative QCD approach to describe the resonant contributions to the $B$ decays into three kaons has been  applied in Ref.~\cite{EPJC80(2020)394}.
As in our case their $B^0 \to K^0 K^+ K^- $ branching ratio is first dominated by the $f_0(980)$ and then by the $\phi(1020)$ contributions.
In their Fig.~3 they show  the different $f_0$ and $f_2^{(')}$ resonance contributions to the $K^+ K^-$ invariant mass distributions but the full spectrum is not calculated and not compared to the existing data.
Quasi-two-body charmless $B$ decays have been recently extensively analyzed in Ref.~\cite{PRD107_116023}  under the factorization-assisted topological-amplitude approach.

In a quasi-two-body QCD factorization (QCDF) framework, the $B^\pm \to K^+ K^- K^\pm$ decays have been studied in Ref.~\cite{PLB699_102}.
The kaon-scalar and vector-form factors describe the strong $K^+K^-$ final state interactions.
A unitary model, which incorporates the scalar $f_0$ resonances, is built for the scalar strange and non-strange kaon form factors.
The vector form factors originate from an existing study on electromagnetic kaon-form factors.
The four parameter fit of this model leads to an overall reasonable agreement with the available Belle and \textit{BABAR} data as can be seen in their fit to some  $K^+K^-$ mass distributions shown in their figures 2 and 3.
In the  $K^+ K^-$-mass spectrum  dominated by the $S$ wave, a  large {\it CP} asymmetry
has been predicted.
These predictions have been confirmed by  {\it BABAR}~\cite{PRD85_112010}  and LHCb~\cite{PRL111_101801}.
With the addition of the $K^+ K^-$-$D$ wave, $f_2(1270)$ resonance, an extension of the just described model~\cite{PLB699_102} is developed by two of the authors in Ref.~\cite{KKK}.
There, the $K^+ K^-$ invariant mass squared dependence of the {\it CP} asymmetry is reproduced in a satisfactory way in the region below~1.9~(GeV)$^2$.

In view of further amplitude analyses, we derive here, also within a  quasi-two-body QCDF framework, the $ B^0 \to K^0_S K^+ K^- $  decay amplitude in terms of kaon form factors and  $B^0$ to two-kaon-transition functions.
These include the resonant and NR parts of the two kaon interactions.
It has been shown, in quantum field theory and using dispersion relations~\cite{Barton1965}, that strong-interaction meson-meson form factors can be calculated exactly provided one knows the meson-meson scattering amplitudes at all energies.
The charmless three-body $B$-meson decays data can also be useful for a better knowledge of the meson-meson strong interactions.
In the  kaon-kaon final state interactions we take into account the $S$, $P$, and $D$  waves. Unitarity is satisfied when the two kaons are in a scalar state.
Here, the final states are the same as in the  $D^0 \to~K^0_SK^+ K^-$ process which has been recently studied in Ref.~\cite{JPD_PRD103}.

A detailed QCDF calculation of the full amplitude, following  the derivation of  the  $B^{\pm}\to \pi^+ \pi^- \pi^{\pm}$ decay amplitudes performed in Ref.~\cite{DedonderPol}, can be done.
This amplitude includes, besides important parts,  Okubo-Zweig-Iizuka (OZI)~\cite{OZI} suppressed terms where an explicit or an implicit $d \bar d$ quark pair appears.
In the present work,  neglecting the OZI terms, we show that the dominant contributions of our amplitude can reproduce, in a reasonable way, the  total branching fraction and the Belle~\cite{PRD82_073011} and \textit{BABAR}~\cite{PRD85_112010} Dalitz-plot projections.
Our model can then be used to build a parametrization which,  in a Dalitz-plot analysis, could be an alternative  to the  commonly applied sum of Breit-Wigner type amplitudes~\cite{PRD96_113003}.

In Sec.~\ref{amplitude} we describe  how, starting from the effective weak decay Hamiltonian, the decay amplitude can be obtained within a quasi-two-body QCDF formulation.
We argue for the choice of the probably important parts which we illustrate by tree and penguin quark Feynman diagrams.
Sec.~\ref{SelectedAmplitudes} gives the explicit expressions of these dominant terms.
Results and discussion of our simultaneous fit of Belle~\cite{PRD82_073011} and \textit{BABAR}~\cite{PRD85_112010}  Collaboration data are presented in Sec.~\ref{results}.
A summary of our model, together with some concluding remarks can be found in Sec.~\ref{conclusions}.
A reminder on formulae for $B^0$-$\bar {B}^0$ mixing and for the time-dependent  asymmetry $A_{{CP}}(t)$ is given in Appendix~\ref{B0B0bmixingtimedep}.

\section{The  \mbox{\boldmath $B^0 \to~K^0_S K^+ K^- $} decay amplitude in QCDF framework}
\label{amplitude}

The amplitude for this charmless-three-body hadronic $B$ meson  decay 
is obtained from the effective weak Hamiltonian~\cite{Ali1998,Beneke:2001ev}
\be \label{Heff}
H_{eff}=\frac{G_F}{\sqrt{2}} \sum_{p=u,c}\lambda_p^{(s)}\, \Big[ C_1 O_1^p + C_2 O_2^p
+\sum_{i=3}^{10} C_i O_i + C_{7\gamma} O_{7\gamma} + C_{8g} O_{8g} \Big] + h.c., 
\ee
where  
\be \label{lambdap}
\lambda_p^{(s)}= V_{pb} V^*_{ps}.
\ee
The  $V_{pp'}\ (p'=b,s)$ are the CKM quark-mixing matrix elements. For the Fermi coupling constant $G_F$ we take the value $1.166379 \times 10^{-5}$ GeV$^{-2}$~\cite{PDG2022}. 
We use the Wolfenstein parameters given in Eq.~(12.26) of Ref.~\cite{PDG2018} which lead to $\lambda_u^{(s)}=(0.2659-i~0.7738)\times10^{-3}$ and $\lambda_c^{(s)}= 0.04105+i~0.6872\times10^{-6}$.
The~$C_i(\mu)$ are the Wilson coefficients for the  four-quark operators  $O_i^{(p)}(\mu)$ at a renormalization scale~$\mu$.
The~$O_{1,2}^{p}$ terms are left-handed current-current operators arising from $W$-boson exchange. The $O_{i=3-10}$ terms are QCD and electroweak penguin operators involving a $W$ boson loop with a $u$ or $c$ quark while $O_{7\gamma}$ and $O_{8g}$ are the electromagnetic and chromomagnetic dipole operators~\cite{Beneke:2001ev}.

The amplitude depends on  the Mandelstam invariants 
  \be  \label{3sa}
 s_{\pm}= m_{\pm}^2 =(p_0+p_\pm)^2,\hspace{0.2cm}  s_{0}=m_0^2=(p_+ + p_-)^2,
\ee
 where $p_0$, $p_+$ and $p_-$ are the four-momenta of the $K_S^0$, $K^+$ and $K^-$ mesons, 
respectively. Energy-momentum conservation implies  
\be \label{pB0}
 p_{B^0}=p_0+p_++p_-,\hspace{0.2cm} 
s_0+s_++s_-=m_{B^0}^2+m_{K^0}^2 + 2 m_K^2,
\ee
where $p_{B^0}$ is the $B^0$ four-momentum and  $m_{B^0}$, $m_{K^0}$ and $m_K$ denote the  $ B^0$, the neutral and charged kaon masses, respectively.  In the following we derive, for the $\bar {B}^0 \to~\bar{K}^0 K^+ K^- $ decay, the contributions of the quasi two-body processes,
\be \label{B0bar}
\bar{B}^0\to [K^+ K^-]_{L}\ \bar{K}^0, \hspace{0.1cm} {\rm and} \hspace{0.1cm} \bar{B}^0\to [\bar{K}^0 K^\pm]_{L}\ K^\mp.
\ee
The final  interacting-kaon pairs, $[K^+ K^-]_L$ and $[\bar K^0 K^\pm]_L$ can be in a scalar, $L=S$, vector, $L=P$ or tensor, $L=D$ states.
The  isospin $I$ of the $[K^+ K^-]_L$ pair can be either~0 or~1, while that of the~$[\bar K^0 K^\pm]_L$ pair is~1.
Then, the possible final quasi-two-body $M_1 M_2$ pairs can be: 
\be
 \label{M1K0K+}
M_{1}^{I=1}(p_0+p_ +) \equiv [\bar K^0 (p_0) K^+(p_+)]_{L}^{I=1}, \hspace{0.2cm} M_{2}(p_-) \equiv K^-(p_-), 
\ee
and
\be \label{M2KK}
M_{1}(p_0) \equiv \bar {K}^0 (p_0),  \hspace{0.2cm}M_{2}^{I=0,1}(p_++p_ -) \equiv [K^+(p_+) K^-(p_-)]_{L}^{I=0,1}.
\ee
The different  isospin 1, ${[\bar K^0 \ K^+]}^{I=1}_{S,P,D}$, and isospin 0 and 1, ${[K^+ K^-]}^{I=0,1}_{S,P,D}$, resonances $R_L^I$ contributing to the meson-meson final state  strong interactions are listed\footnote{Beyond this table, the isospin 1 of the $\bar K^0 K^{+}$ states will not be specified unless necessary.} in Table~\ref{Tabfsi}.

\begin{table}[h]
\caption{Two-body resonances  $R_L^I$ contributing, in the $\bar{B}^0\to \bar{K}^0 K^+  K^-$  decays, to the isospin 1 ${[\bar K^0 \ K^+]}^{I=1}_{S,P,D}$, and to the isospin 0 and 1 $[K^+ K^-]^{I=0,1}_{S,P,D}$  final state meson-meson strong interactions.
Our model amplitude does not include the contribution of the $f_2'(1525)$.
The resonances $a_0(980)^0$, $a_0(1450)^0$, $f_2(1270)$ and $a_2(1320)^0$ contribute only to the OZI suppressed parts which we will neglect.
 }
\label{Tabfsi}
\begin{center}
\begin{tabular}{cccc}
\hline
\hline
 Final state & $L=S$ & $L=P$ & $L=D$\\
\hline
$[\bar K^0 K^+ ]^{I=1}_L $ & \ \ $a_0(980)^{+}$, $a_0(1450)^{+}$ &\  \ $\rho(770)^{+}$, $\rho(1450)^{+}$, $\rho(1700)^{+}$&\  \ $a_2(1320)^{+}$ \\
$[K^+ \ K^-]^{I=0}_L $ & $f_0(980)$, $f_0(1370)$, $f_0(1500)$ &\ \  $\omega(782)$, $\omega(1420)$, $\omega(1650)$, $\phi(1020)$, $\phi(1680)$ & $f_2(1270)$\\
$[K^+ \ K^-]^{I=1}_L $ & $a_0(980)^0$, $a_0(1450)^0$ & $\rho(770)^{0}$, $\rho(1450)^{0}$, $\rho(1700)^{0}$& $a_2(1320)^{0}$ \\
\hline
\hline
\end{tabular}
\end{center}
\end{table}

Applying the quasi-two-body QCDF~\cite{Beneke:2001ev} formalism for the  $\bar{B}^0\to {K}^0_S K^+ K^- $ decay and neglecting small {\it CP} violation effects in $K^0_S$ decays by using
\be \label{K0S}
\vert K^0_S \rangle \approx \frac{1}{\sqrt{2}}\  \left (\vert K^0 \rangle+ \vert \bar{K}^0 \rangle \right ),
\ee
the matrix elements of the effective weak Hamiltonian~(\ref{Heff}) can be written as (see Eqs.~(2.1) and (A1) of Ref.~\cite{ChengPRD76_094006})
\bqa \label{BK0KKTp}
{\bar A}(s_{0},s_{-},s_{+})& \equiv &  \frac{1}{\sqrt{2}} \left \langle \bar{K}^0(p_0) K^+(p_+)\ {K^-}(p_-) \vert H_{eff}\vert \bar{B}^0(p_{B^0}) \right \rangle  \nonumber \\
& = &  
\frac{G_F}{ {{2}}}\ \left \{ \lambda_u^{(s)} \ \left \langle \bar{K}^0 K^+K^- \vert T_u \vert \bar{B}^0\right \rangle
+ \lambda_c^{(s)} \ \left \langle \bar{K}^0 K^+K^- \vert T_c \vert \bar{B}^0\right \rangle \right \},
\eqa
with
\bqa \label{Tp}
\left \langle \bar{K}^0 K^+ K^-\vert T_p\vert \bar{B}^0\right \rangle&=&  \langle   \bar{K}^0 K^+ K^-\vert \Big\{ a_1\ \delta_{pu}\ (\bar u b)_{V-A} \otimes (\bar s u)_{V-A}\nonumber\\
	&+&  a_2\ \delta_{pu}\ (\bar s b)_{V-A} \otimes (\bar u u)_{V-A} 
	+a_3\ (\bar s b)_{V-A} \otimes \sum_q (\bar q q)_{V-A} \nonumber\\
	&+&a_4^p\sum_q(\bar q b)_{V-A} \otimes (\bar s q)_{V-A} 
	+a_5\ (\bar s b)_{V-A} \otimes \sum_q(\bar  q q)_{V+A} \nonumber\\
	&-& 2 \ a_6^p\sum_q (\bar q b)_{sc-ps} \otimes (\bar s q)_{sc+ps} 
	+a_7\ (\bar s b)_{V-A} \otimes \sum_q\frac{3}{2}e_q (\bar q q)_{V+A} \nonumber\\
	&-& 2\  a_8^p\sum_q(\bar q b)_{sc-ps} \otimes \frac{3}{2}e_q (\bar s q)_{sc+ps}
	+a_{9}\  (\bar s b)_{V-A} \otimes \sum_q\frac{3}{2}e_q (\bar q q)_{V-A}\nonumber\\
	&+&a_{10}\sum_q(\bar q b)_{V-A} \otimes \frac{3}{2}e_q (\bar s q)_{V-A}\Big\} \vert \bar B^0 \rangle,
\eqa
where $p=u$ or $c$ and $a_j^{(p)}$ are effective QCDF coefficients.
For simplicity, in Eq.~(\ref{Tp}) we have not specified their argument $(M_1M_2)$. These $a_j^{(p)}(M_1M_2$) coefficients\footnote{In the following, as done in Eq.~(\ref{Tp}), these arguments $M_1M_2$ will not be specified, unless necessary.} are asymmetric in $M_1\leftrightarrow M_2$ with $M_2$ relevant for short distance dynamics as the final meson $M_2$ denotes the meson which does not include the spectator $\bar d$ quark  of the $\bar B^0$.
This implies that the meson $M_1$ is either the $\bar K^0$ itself or contains it [see Eqs.~(\ref{M1K0K+}) and~(\ref{M2KK})].
In Eq.~(\ref{Tp}),  $(\bar q_{1}q_{2})_{V \mp A} = \bar q_{1} \gamma_\mu(1\mp \gamma_5)q_{2}$, $(\bar q_1 q_2)_{sc\pm ps}=\bar q_1(1\pm \gamma_5)q_2$ and $e_q$ denotes the electric charge of the quark $q$ in  units of the elementary charge $e$. 
The sum on the index $q$ runs over $u$, $d$, $s$ and the summation over the color degree of freedom has been performed. 
The notations $sc$ and $ps$ stand for scalar and pseudoscalar, respectively.
The symbol $\otimes$ indicates that the different components of the matrix elements are to be calculated in the factorized form.
The $[K^+ K^-]_L$ states are assumed to originate from a $u\bar{u}$ or $s\bar{s}$ or $d\bar{d}$ pair and  the $[\bar K^0 K^{+}]_L$   states from a $\bar{d}u $ one. 

The $a_j^p$ quantities, at next-to-leading order (NLO) in the strong coupling constant $\alpha_s$, can be written in terms of the Wilson coefficients as~\cite{bene03}
 \be \label{ajp}
 a_j^{(p)}(M_1M_2)=\left(C_j+\frac{C_{j\pm 1}}{N_C}\right) N_j(M_2)+ \frac{C_{j\pm 1}}{N_C}\frac{C_F \ \alpha_s}{4\pi}
 \left [V_j(M_2)+\frac{4\pi^2}{N_C}H_j(M_1M_2)\right] +P_j^p(M_2),
 \ee
 where the upper (lower) signs apply when the index $j$ is odd (even), $N_C=3$ is the number of colors and $C_F=(N_C^2-1)/2N_C$.
Note that in the leading-order (LO) contribution $N_j(M_2)=0$ for $M_2= [K^+K^-]_{P}$ and $j=6, 8$, otherwise  $N_j(M_2)=1$.
The NLO quantities $V_j(M_2)$ come from one-loop vertex corrections, $H_j(M_1M_2)$ from hard spectator scattering interactions and $P_j^p(M_2)$ from penguin contractions.
For $ j=1$, $2$, $3$, $5$, $7$ and $9$, the superscript $p$ in $a_j^{(p)}(M_1M_2)$ is to be omitted since the penguin corrections are equal to zero in these cases.
The NLO hard scattering corrections require the introduction of four phenomenological parameters to regularize end point divergences  related to asymptotic wave functions~\cite{bene03}. 

From Eqs.~(\ref{BK0KKTp}) and (\ref{Tp}) one can write the full factorized $\bar B^0 \to {K}_S^0 K^+ K^- $ amplitude, $\bar{A}(s_{0},s_{-},s_{+})$, as (see\footnote{\label{ChengOZI}Following Ref.~\cite {ChengPRD76_094006},  we keep terms with intermediate $d$ and $\bar d$  quarks in the factorized amplitude.}  Eq.~(A4) of Ref.~\cite{ChengPRD76_094006}) 
\be  \label{Mfull} 
{\bar A}(s_{0},s_{-},s_{+}) = \sum_{i=1}^{9} \sum_{L=S,P,D} \sum_{I=0,1} {\bar {\cal{A}}}_{iL,I}(s_{0},s_{-},s_{+}) =  \frac{G_F}{2} \sum_{i=1}^{ 9} \sum_{p=u,c}  \lambda_p^{(s)} {\cal{H}}^{(p)}_i 
\ee
with
 \begin{eqnarray}  \label{TpK0KK}
  {\cal{H}}^{(p)}_1 & = & \langle\bar K^0 K^+  |(\bar u b)_{V-A}|\bar B^0\rangle \cdot \langle K^-|(\bar s u)_{V-A}|0\rangle
 \left[a_1 \delta_{pu}+a^p_4+a_{10}^p-(a^p_6+a^p_8) r_\chi^K\right] \nonumber\\
{\cal{H}}^{(p)}_2 &= & \langle K^+  K^-|(\bar d b)_{V-A}|\bar B^0\rangle \cdot \langle \bar K^0|(\bar s d)_{V-A}|0\rangle
 (a^p_4-{1\over 2}a^p_{10})  \nonumber\\
{ \cal{H}}^{(p)}_3 & = &\langle \bar K^0|(\bar s b)_{V-A}|\bar B ^0\rangle \cdot
                                    \langle K^+ K^-|(\bar u u)_{V-A}|0\rangle
    (a_2\delta_{pu}+a_3+a_5+a_7+a_9)    \nonumber\\
{ \cal{H}}^{(p)}_4 & = &\langle \bar K^0|(\bar s b)_{V-A}|\bar B ^0\rangle \cdot  \langle K^+ K^-|(\bar d d)_{V-A}|0\rangle
    \left[a_3+a_5-\frac{1}{2}(a_7+a_9)\right]  \nonumber\\
 { \cal{H}}^{(p)}_5 & = &\langle \bar K^0|(\bar s b)_{V-A}|\bar B ^0\rangle \cdot  \langle K^+ K^-|(\bar s s)_{V-A}|0\rangle
    \left[a_3+a^p_4+a_5-\frac{1}{2}(a_7+a_9+a^p_{10})\right]  \nonumber\\
 { \cal{H}}^{(p)}_6 & = & \langle \bar K^0|(\bar s b)_{sc}|\bar B^0\rangle
       \langle K^+K^-|(\bar s s)_{sc}|0\rangle (-2 a^p_6+a^p_8)  \nonumber\\
 { \cal{H}}^{(p)}_7 & = &\langle K^+ K^-|(\bar d b)_{sc-ps}|\bar B ^0\rangle  \langle \bar K^0|(\bar sd)_{sc+ps}|0\rangle
   (-2a_6^p+a_8^p)  \nonumber\\
{ \cal{H}}^{(p)}_8 & = &\langle \bar K^0 K^+ K^-|(\bar s d)_{V-A}|0\rangle \cdot \langle 0|(\bar d b)_{V-A}|\bar B^0\rangle
  \left(a^p_4-{1\over 2}a^p_{10}\right) \nonumber\\
{ \cal{H}}^{(p)}_9 & = & \langle \bar K^0 K^+ K^-|(\bar s d)_{ps}|0\rangle \langle 0|(\bar d b)_{ps}|\bar B^0\rangle (-2a^p_6+a^p_8).
 \end{eqnarray}
The chiral factor $r_\chi^K$ is given by  $r_\chi^K= 2m_K^2/[(m_b+m_d)(m_u+m_s)]$, $m_b$, $m_d$, $m_u$ and $m_s$ being the $b$-, $d$-, $u$- and $s$-quark masses, respectively and $p=u$ or $c$.
Because the isospin of the $s$ quark is 0, the $\bar s s$ pair  in $ {\cal{H}}^{(p)}_5$ and $ { \cal{H}}^{(p)}_6$ generates only isospin 0 states.
\begin{figure}[h]  \begin{center}
\includegraphics[scale=0.345]{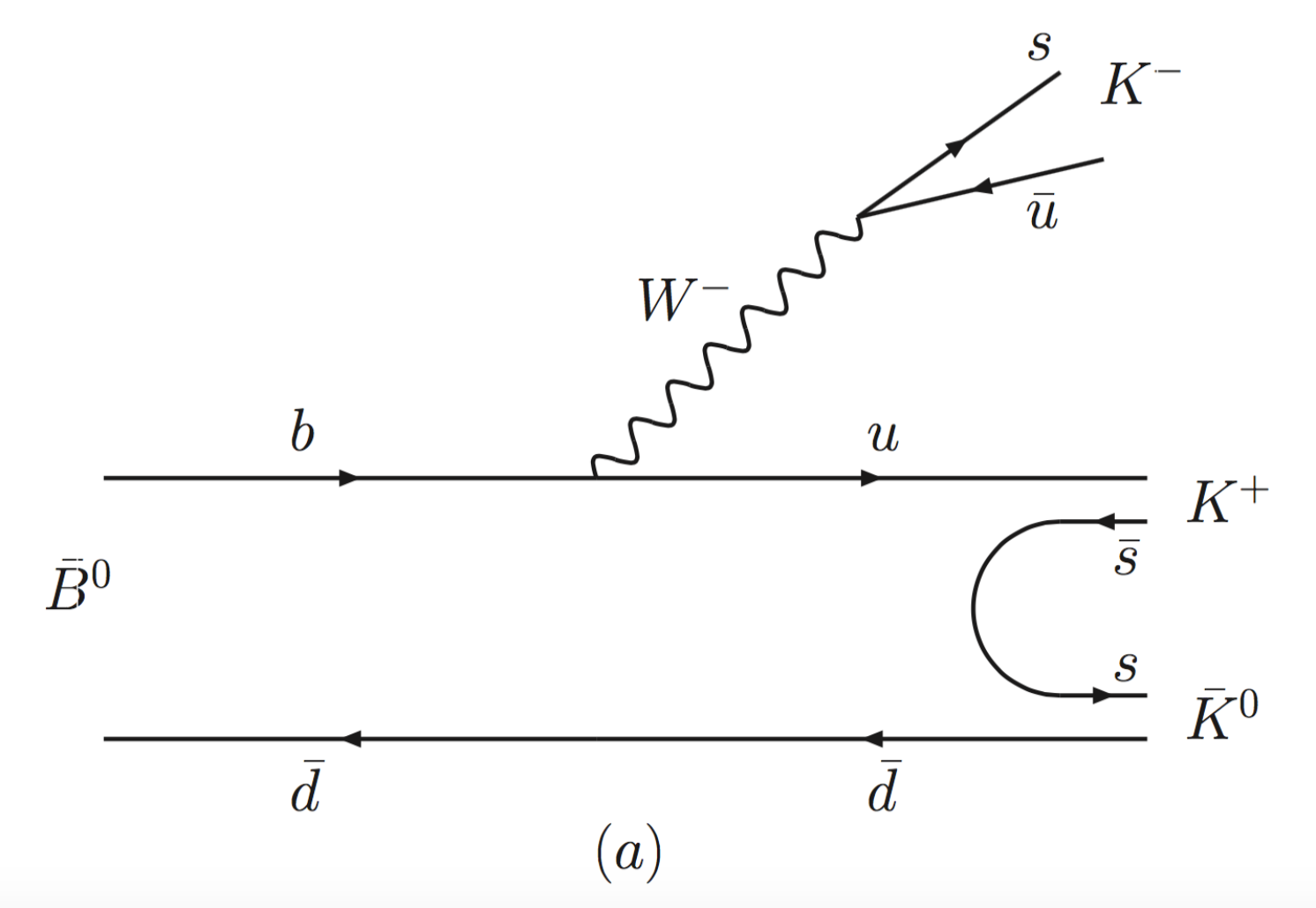}
\includegraphics[scale=0.345]{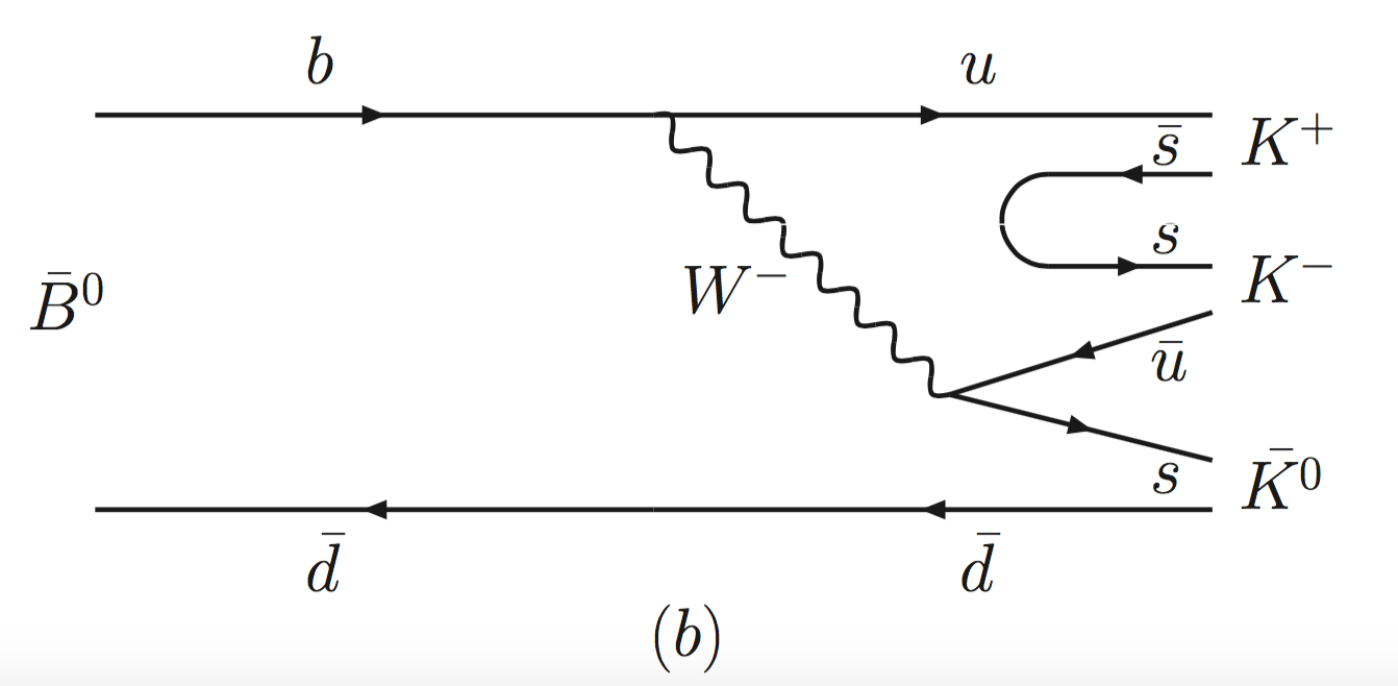}
\caption{Quark Feynman tree diagrams for the decay  $\bar B^0 \to \bar{K}^0 K^+ K^- $: 
(a) for the color favored  ${\cal{H}}^{(p)}_1$  term proportional to $a_1$ and (b) for the color suppressed  ${\cal{H}}^{(p)}_3$  term  proportional to $a_2$.
}\label{F1}
\end{center} \end{figure}

Inspection of the ${\cal{H}}^{(p)}_i$ in Eqs.~(\ref{TpK0KK}) tells us that some of them are expected to make a fairly small contribution to the  $\bar B^0 \to \bar{K}^0K^+ K^-$ amplitude.
In  ${\cal{H}}^{(p)}_4$ the  formation of the final state $K^+ K^-$ goes through an explicit $d \bar d$ pair.
In the $i=2$ and $i=7$ to~9~terms, this creation results from an implicit $d \bar d$ pair due to the presence of a $d$ and $\bar d$ quarks in their matrix elements.
These terms lead naturally to $K^0 \bar K^0$ production and they require a supplementary final state interaction to produce a $K^+ K^-$ pair.
At the microscopic level a $d \bar d$ quark annihilation followed by $s \bar s$ and $u \bar u$ pair creation can only be depicted by non-planar quark diagrams which give small contributions to the decay amplitude.
Furthermore, as can be seen in Table~1 of Ref.~\cite{PLB699_102}, the  NLO effective Wilson coefficients $a_j^{(p)}$ for $j>2$ are small and those for $j>6$  smaller.
For $j>1$ their real part is only few percent of  that of $a_1$. Accordingly, we do not calculate the parts corresponding to these OZI suppressed matrix elements, ${\cal{H}}^{(p)}_2$, ${\cal{H}}^{(p)}_4$, ${\cal{H}}^{(p)}_7$ and  ${\cal{H}}^{(p)}_{8,9}$ ($\bar {B}^0$ annihilation terms).

\begin{figure}[h]  \begin{center}
\includegraphics[scale=0.36]{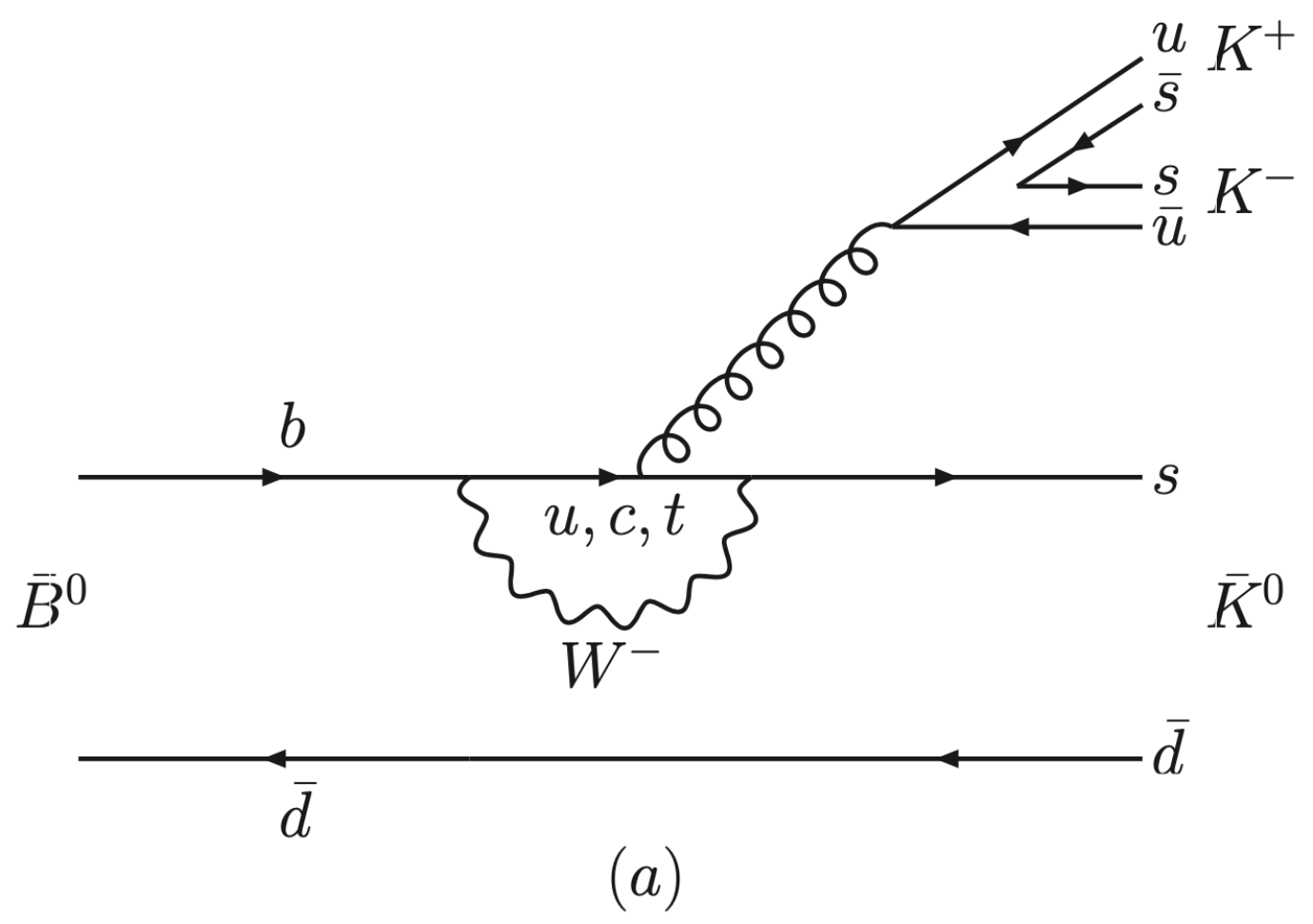}
\includegraphics[scale=0.36]{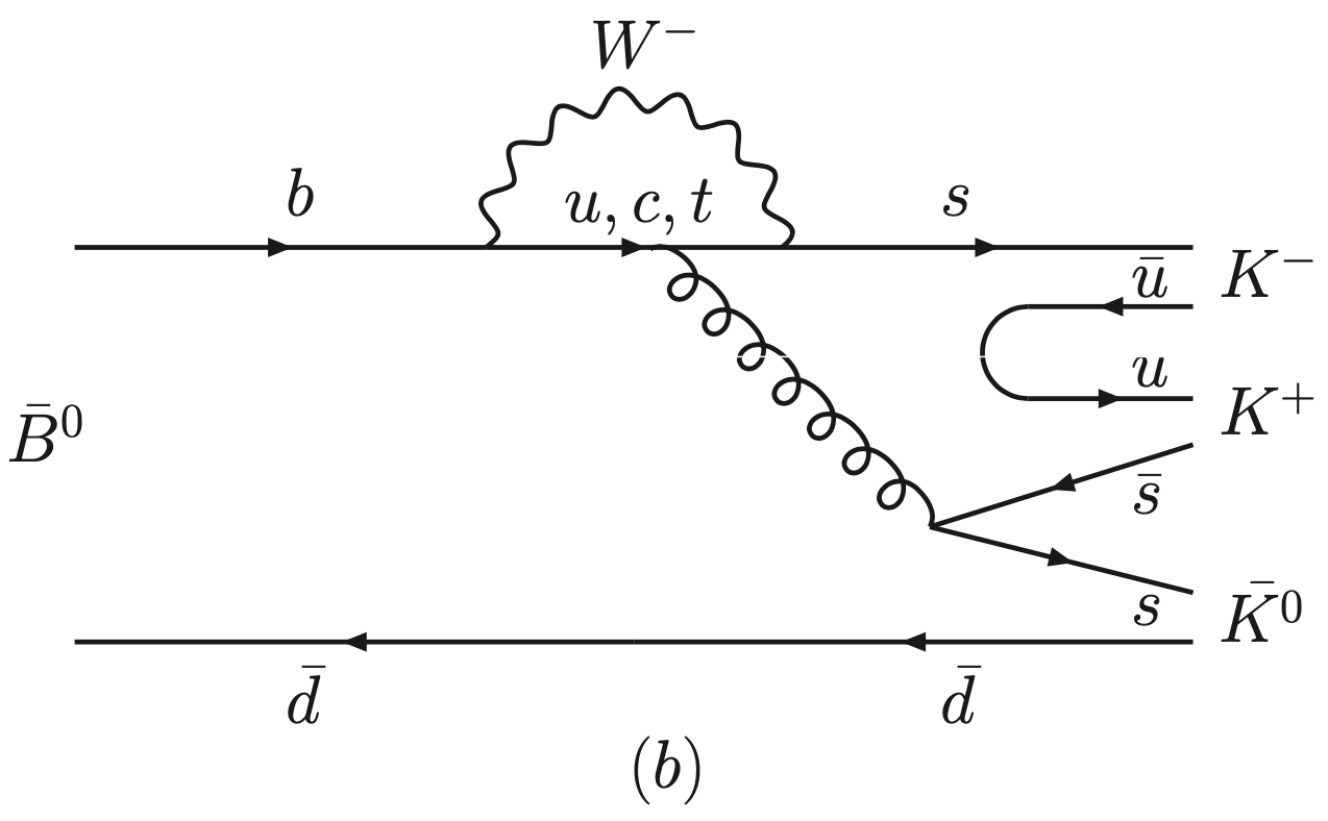}
\caption{Quark Feynman penguin diagrams for the decay  $\bar B^0 \to \bar{K}^0 K^+ K^- $: (a) for the ${\cal{H}}^{(p)}_3$ term and (b) for the ${\cal{H}}^{(p)}_5$ and  ${\cal{H}}^{(p)}_6$ terms.
The effective gluon exchange is represented by a spring like line.
}\label{F2}
\end{center} \end{figure}

\begin{figure}[h]  \begin{center}
\includegraphics[scale=0.45]{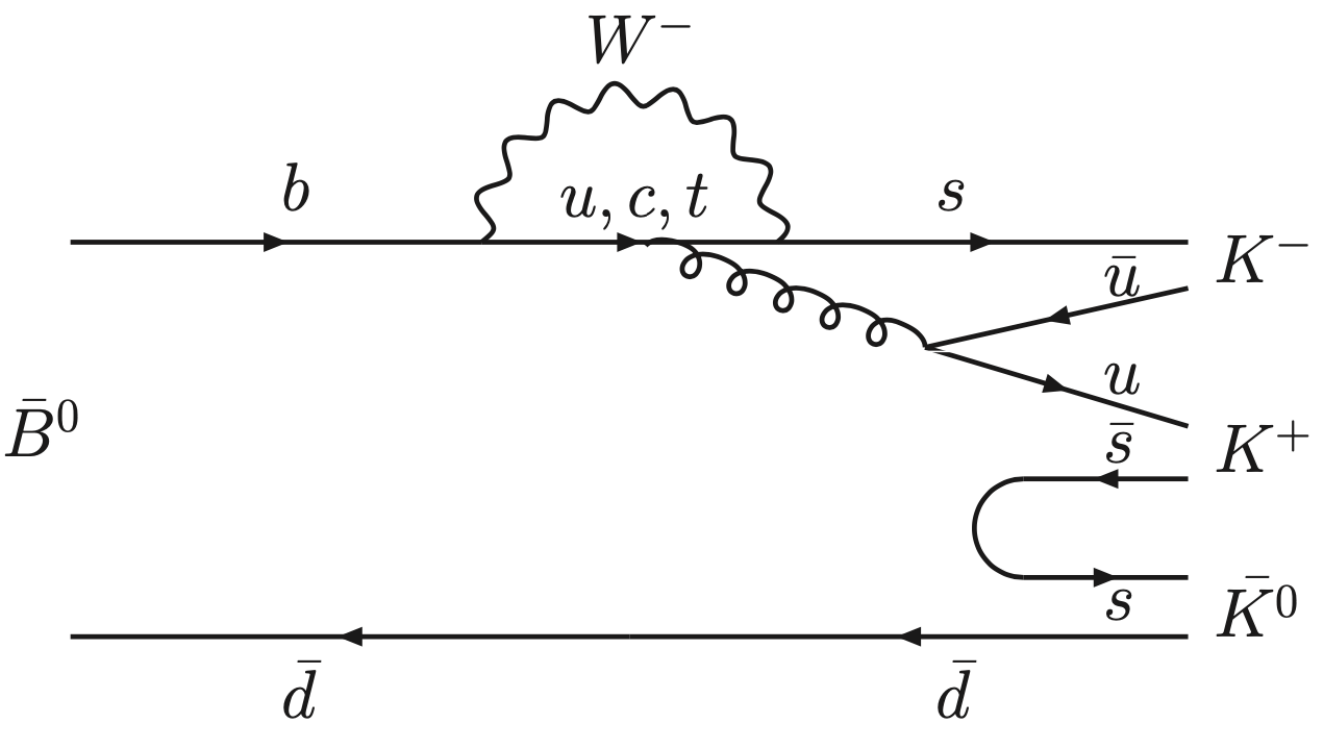}
\caption{ As in Fig.~\ref{F2} but for the  ${\cal{H}}^{(p)}_1$ term.
}\label{F3}
\end{center} \end{figure}

One expects large contributions to the amplitude from i)  ${\cal{H}}^{(p)}_1$, the Wilson coefficient $a_1$ being the dominant one (see Table~1 of Ref.~\cite{PLB699_102}) and from ii) ${\cal{H}}^{(p)}_{3,5,6}$ because these terms are proportional  to the kaon form factors.
The quark processes involved in these terms can be represented by the Feynman diagrams depicted in Figs.~\ref{F1} to~\ref{F3}. 
The wavy lines stand for $W^\pm$ exchanges, the spring-like lines, if any, for a gluon and the straight lines with an arrow pointing to the right (left) for a quark (antiquark).
The short distance $a_1$ contribution of ${\cal{H}}^{(p)}_1$ corresponds to the color favored tree diagram shown in Fig.~\ref{F1}(a).
The color suppressed $a_2$ term of ${\cal{H}}^{(p)}_3$ arises from the tree diagram drawn in Fig.~\ref{F1}(b).
The {$a_j^{(p)}, j>2$} contributions of   ${\cal{H}}^{(p)}_3$, ${\cal{H}}^{(p)}_5$  and  ${\cal{H}}^{(p)}_6$ can be represented by the penguin diagrams of   Fig.~\ref{F2} and that of ${\cal{H}}^{(p)}_1$ by the penguin diagram of Fig.~\ref{F3}. 
The factorized forms given in  Eqs.~(\ref{TpK0KK}) can be understood if, in the diagrams of Figs.~\ref{F1} to~\ref{F3} one replaces the very heavy $W$ meson exchange by a vacuum state creation.

In a way similar to that developed in Ref.~\cite{DedonderPol} for the $B^{\pm}\to \pi^+ \pi^- \pi^{\pm}$ decays, the detailed expressions for the different ${\bar {\cal{A}}}_{iL,I}(s_{0},s_{-},s_{+})$ amplitudes which build up the ${\cal{H}}^{(p)}_i$ contributions can be given as product of short distance terms [sum of $a_j^{(p)}(M_1M_2)$] by long distance ones which can be expressed or are given in terms of meson-meson form factors.
As mentioned in the previous paragraph, the amplitudes coming from the terms  ${\cal{H}}^{(p)}_3$, ${\cal{H}}^{(p)}_5$  and  ${\cal{H}}^{(p)}_6$ are directly proportional to the kaon form factors.
For  ${\cal{H}}^{(p)}_1$ one has to evaluate  the matrix elements of $\bar B^0$ transitions to two-kaon states.
As in the previous studies~\cite{PRD96_113003}, assuming this transition to proceed through the dominant intermediate resonances, it can be approximated, either by a phenomenological function calculated via a unitary equation, or as being proportional to the  isovector kaon form factors.
In the calculation of the scalar product of two matrix elements in Eqs.~(\ref{TpK0KK}) one makes use of Eqs.~(B1) and (B6) of Ref.~\cite{ChengPRD76_094006}.
As argued above, only the important parts of the amplitude, needed to reasonably reproduce the currently available experimental total branching fraction and the  Belle~\cite{PRD82_073011} and \textit{BABAR}~\cite{PRD85_112010} Dalitz plot projections, are given in next Section.

\begin{table*}[h]
\caption{Values  of the different decay constants (in GeV) and of the fixed form factors used in our model.}
\label{fixed}
\begin{center}
\begin{tabular}{lcc}
\hline
\hline
\hspace{0.0cm}Parameter &Value & Reference\\ 
\hline
$f_{K^+}=f_{K^-}\equiv f_K$          &   0.1561  &  \cite{PDG2022}   \\ 
$f_{\rho^+}=f_{\rho^-}\equiv f_{\rho}$                 &   0.209   &  \cite{bene03}      \\
$F_0^{\bar{B^0} a_0^+}(m_{K}^2)=\sqrt{2} F_0^{\bar{B^0} a_0^0}(m_{K}^2)=\sqrt{2} F_0^{\bar{B^0} f_0}(m_{K}^2)$  & 0.18 & \cite{El-Bennich_PRD79}    \\
$A_0^{B^0 \rho^+}(m_{K}^2)=\sqrt{2} A_0^{B^0 \rho^0}(m_{K}^2)$  &   0.52   & \cite{bene03} \\
$F^{\bar{B^0} a_2^+}(m_{K}^2, m_{a_2}^2)=\sqrt{2} F^{\bar{B^0} a_2^0}(m_{K}^2, m_{a_2}^2)$& 0.14 &   \cite{KimPRD67}            \\
\hline
\hline
\end{tabular}
\end{center}
\end{table*}

\section{Dominant  contributions to the amplitude}
\label{SelectedAmplitudes}

 We will give the dominant parts of the $\bar {B}^0 \to~\bar{K}^0 K^+ K^- $ decay amplitude and, applying charge conjugation transformation, the corresponding $B^0 \to~{K}^0 K^- K^+$ ones.
 Within this transfomation, the final $K^\pm$ mesons will be exchanged with the $K^\mp$ ones and the $s_\pm$ Mandelstam invariants with the $s_\mp$ ones.
The decay constants and the fixed form-factor values entering our model are given in Table~\ref{fixed}.
The values for the quark and meson masses are listed in Table~\ref{masses}.
For the parts of the amplitude arising from the  ${\cal{H}}^{(p)}_1$  term [see Eqs.~(\ref{TpK0KK})] which involve  the calculation of the $\bar B^0$ transition to two kaons,  viz. $\langle\bar K^0 K^+  |(\bar u b)_{V-A}|\bar B^0\rangle$ our derivation will follow partly that reported in appendix A of Ref.~\cite{DedonderPol} for the $\langle\pi^+ \pi^-  |(\bar u b)_{V-A}|\bar B^-\rangle$ matrix element completed by the use of an equation similar to Eq.~(20) of Ref.~\cite{JPD_PRD103}.

As seen in the previous Section the different contributions to the amplitude are proportional to the sums of the effective Wilson coefficients\footnote{As pointed out in the paragraph below Eq.~(\ref{Tp}) the meson position in the $M_1M_2$ pair matters.} $a_j^{(p)}(M_1 M_2)$~(\ref{ajp}).
We show below that these sums are given by the functions $\bar \nu $, $\bar y $, $\bar w_u$, and $ \bar w_s$ [see Eqs.~(\ref{vuc}), (\ref{RSpi}), (\ref{wu}) and (\ref{ws})].
Following Ref.~\cite{PLB699_102}, for the calculation of the Wilson coefficients, we take into account one-loop vertex and penguin corrections but neglect hard scatering ones.
Then  one has $a_j^{(p)}(\bar{K}^0 R_{P})\equiv a_{jw}^{(p)}$,
 $a_j^{(p)}(\bar{K}^0 R_{S})\equiv a_{j\nu}^{(p)}$ and
$a_j^{(p)}(R_{S} M_2)= a_j^{(p)}(R_{P} M_2)=a_j^{(p)}(R_{D} M_2) \equiv a_{jy}^{(p)}$.
We use the corresponding NLO values calculated and given in Ref.~\cite{PLB699_102}.
These are evaluated at the renormalization scale $\mu=m_b/2$~\cite{bene03}.

  \begin{table*}[h]
\caption{Values of the different quark and meson masses (in GeV)~\cite{PDG2022}  entering our model amplitude.}
\label{masses}
\begin{center}
\begin{tabular}{lcccc}
\hline
\hline
$m_u$   & $m_d$  &   $m_s$  &   $m_b$ \\
\hline
  0.0022 & 0.0047 & 0.095  & 4.18 \\   
\hline
 $m_{\pi^\pm}$   &   $m_{K^0}$   &  $m_{K^\pm}$   &   $m_{B^0}$ \\
\hline
  0.139570    &  0.497611 &    0.493677    &   5.27963   \\ 
\hline
\hline
\end{tabular}
\end{center}
\end{table*}

\begin{figure}[h]  \begin{center}
\includegraphics[scale=0.65]{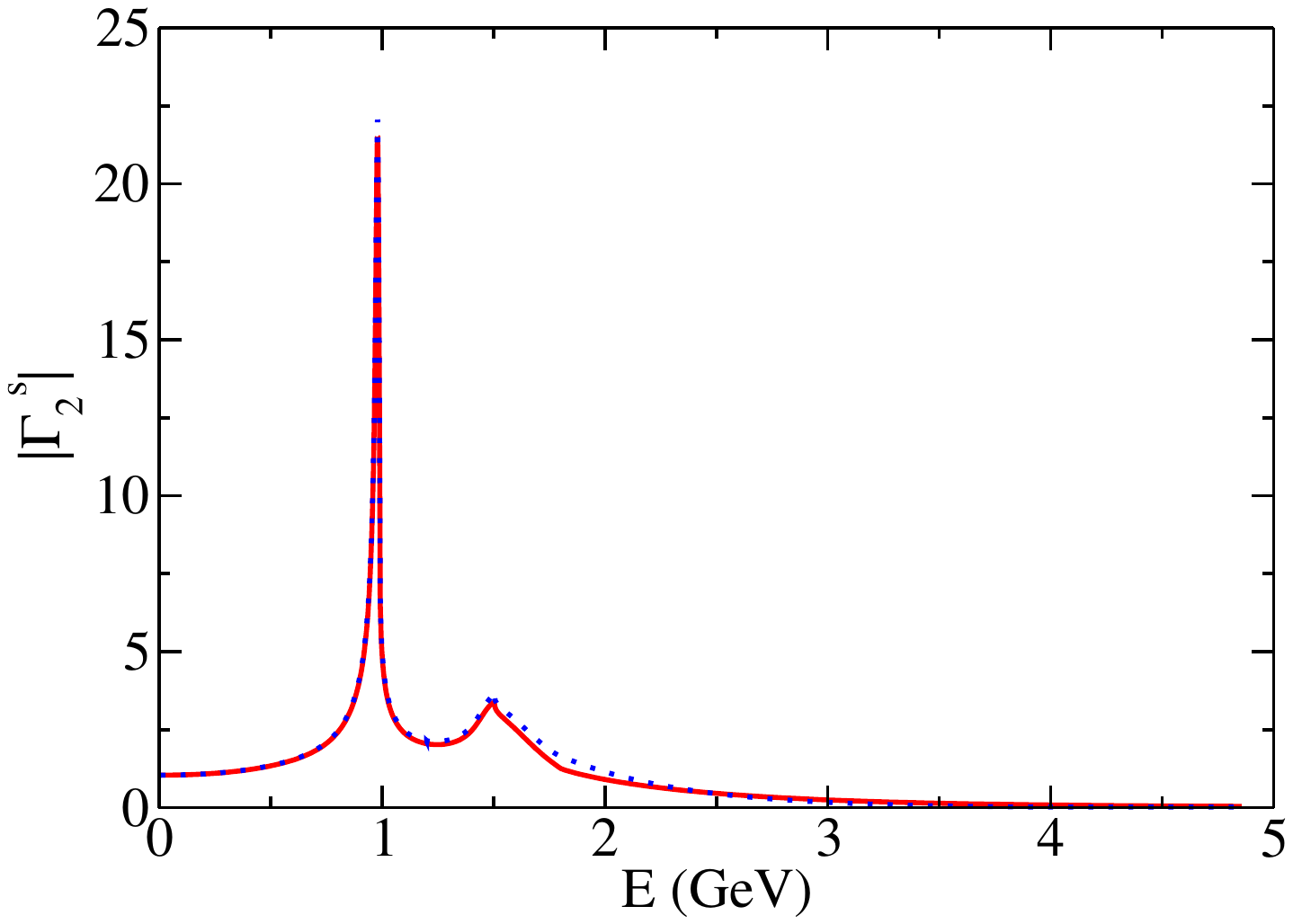}
\caption{Modulus, $|\Gamma_2^s|$, of the strange scalar-isoscalar  kaon form factor  
$\Gamma_2^{s}(s_0)$  ($E=\sqrt{s_0}$) calculated~\cite{Moussallam_2000, Moussallam_2020}   in the dispersion-relation framework using  the updated~\cite{JPD_PRD103} $T$ matrix  of the  $\pi \pi$ (channel 1), $K \bar K $ (channel 2) and effective $(2\pi)(2\pi)$ (channel 3) coupled-channel model of Ref.~\cite{EPJ}. 
Solid (red) line: calculation done with the asymptotic phase shift $\delta_{11} (s_0 \to \infty)= 2 \pi$, $\delta_{22} (s_0 \to \infty)$ = 0 and  $\delta_{33} (s_0 \to \infty) = \pi$.
For the dot (blue) line: $\delta_{11} (s_0 \to \infty)= 2 \pi$, $\delta_{22} (s_0 \to \infty) =\pi$ and  $\delta_{33} (s_0 \to \infty) = 0$.}
\label{ModGam2s}
\end{center} \end{figure}

 \subsection{Contributions to the amplitude with two kaons in $S$ wave}
\label{2kaonSwav}
\subsubsection{The $K^+ K^-$ contribution}
We retain the part coming from the ${\cal{H}}^{(p)}_6$  term in Eqs.~(\ref{TpK0KK}) where the final $K^+ K^-$ forms a scalar and  isoscalar state [see Fig.~2(b)].
We have for this $\bar{B^0}\to K^0_S K^+ K^-$ term
\be \label{M6SI}
\bar{A}_1(s_{0},s_{-},s_{+})\equiv {\bar {\cal{A}}}_{6S,0}(s_{0},s_{-},s_{+}) =  G_F\ \bar \nu(\bar K^0 f_0) \langle \bar K^0|(\bar s  b)_{sc}|\bar B^0\rangle  \langle K^+K^-|(\bar s s)_{sc}|0\rangle, 
\ee
with [see Eqs.~(\ref{Mfull}), (\ref{TpK0KK}) and also  Eq.~(11) in Ref.~\cite{PLB699_102}]
\be \label{vuc}
\bar \nu = \lambda_u^{(s)} \left (- a^u_{6\nu} + \frac{1}{2}\ a^u_{8\nu}\right )
+ \lambda_c^{(s)} \left (- a^c_{6\nu} + \frac{1}{2}\ a^c_{8\nu}\right ).
\ee
The intermediate scalar-isoscalar $K^+ K^-$ resonances for invariant $m_0$ masses $\lesssim$~1.6~GeV~\cite{JPD_PRD103} correspond to the $f_0$ family, mainly $f_0(980)$, $f_0(1370)$ and $f_0(1500)$ which we denote as $f_0$.
Using the $s$ and $b$ quark equations of motion and Eq.~(B6) of Ref.~\cite {ChengPRD76_094006} one gets 
\be
\langle \bar K^0|(\bar s  b)_{sc}|\bar B^0\rangle = \frac{m^2_{B^0}- m^2_K}{m_b - m_s} \ F_0^{\bar B^0 \bar K^0}(s_0).
\ee
For the $\bar B^0$ to $\bar K^0$ transition form factor, we take~\cite{Ball_PRD71_014015}
\be \label{F0B0K0}
 F_0^{\bar B^0 \bar K^0}(s) =  \frac{r_0}{1-\frac{s}{s_t}},
\ee
where $r_0 = 0.33$ and $s_t= 37.46$ GeV$^2$. One introduces (Eq.~(10) of Ref.~\cite{fkll}) the strange form factor  $\Gamma_2^{s}(s_0)$ with
\be \label{gamma2s} 
 \langle K^+(p_+)K^-(p_-)|\bar s s|0 \rangle =   \ B_0\ \Gamma_2^{s*}(s_0).
 \ee
The quantity $B_0$ is related to the vacuum quark condensate, as in Ref.~\cite{fkll} we use
\be\label{B0}
 B_0=\frac{m_\pi^2}{m_u+m_d},
\ee 
where $m_\pi$ is the charged pion mass.
Then we obtain the following contribution for the  $\bar B^0$  case,
\be
\label{A6S0b}
\bar{A}_1(s_{0},s_{-},s_{+}) =  G_F\ \bar{\nu}(\bar K^0 f_0) \ \frac{m_{B^0}^2-m_{K^0}^2}{m_b-m_s}\ {B_0} \ \Gamma_2^{s*}(s_0) \ F_0^{\bar B^0 \bar K^0}(s_0).
\ee
For the $B^0$ we have
\be
\label{A6S0}
{A}_1(s_{0},s_{+},s_{-}) =  G_F\ {\nu}( K^0 f_0) \ \frac{m_{B^0}^2-m_{K^0}^2}{m_b-m_s}\ {B_0} \ 
\Gamma_2^{s*}(s_0) \ F_0^{ B^0 K^0}(s_0),
\ee
with, from charge conjugation symmetry,  $F_0^{ B^0 K^0}(s_0)=F_0^{\bar B^0 \bar K^0}(s_0)$ and  ${\nu}( K^0 f_0)=  \bar{\nu}(\bar K^0 f_0; \lambda_p^{(s)}\to\lambda_p^{(s)* }|_{p=u,c} )$.

The form factor  $\Gamma_2^{s}(s_0)$ has been caculated by B.~Moussallam~\cite{Moussallam_2000, Moussallam_2020} in the Muskhelishvili-Omn\`es (MO) dispersion-relation framework~\cite{Barton1965, MO}. 
B. Moussallam has used the updated  $S$~matrix  of the  $\pi \pi$ (channel~1), $K \bar K$ (channel~2) and effective $(2\pi)(2\pi)$ (channel~3) coupled-channel model of Ref.~\cite{EPJ}.
Details on this scattering $S$ matrix can be found in Appendix~A of Ref.~\cite{JPD_PRD103}.
As can be seen in Fig.~\ref{ModGam2s}  the modulus of $\Gamma_2^{s*}(s_0)$ ($E=\sqrt{s_0}$) 
has a $K^+K^-$ threshold peak which is due to the $f_0(980)$ resonance. 
The bump near 1.5~GeV arises from the opening of the third effective $4 \pi$ channel close to  $2  m_\rho$ where $m_\rho$ is the $\rho(770)$ mass~\cite{JPD_PRD103, EPJ}.
Here, the $S$ matrix has several poles located nearby and these have an important influence on the energy behavior of $\Gamma_2^{s}(s_0)$ in this region.
These poles could be related to the $f_0(1370)$ and $f_0(1500)$ resonances.
In our model, we use the form factor corresponding to the red solid line of Fig.~\ref{ModGam2s} where $\delta_{11} (s_0 \to \infty)$, $\delta_{22} (s_0 \to \infty)$, $\delta_{33} (s_0 \to \infty)$   equal $2\pi, 0$ and $\pi$, respectively.

\subsubsection{The  $ K^0_S K^+$ contribution}

 As seen from Eqs.~(\ref{Mfull}) and~(\ref{TpK0KK}), the ${\cal{H}}^{(p)}_1$ contribution gives rise to the part ${\bar {\cal{A}}}_{1S,1}$  with the $\bar K^0 K^+$ pair in a  scalar-isovector  state (see Figs.~1(a) and~3).
One has, 
\be
\label{M1L1}
\bar{A}_2(s_{0},s_{-},s_{+})\equiv {\bar {\cal{A}}}_{1S,1}(s_{0},s_{-},s_{+})=\frac{G_F}{2}\ \bar y(R_S K^-) \left\langle [\bar K^0 K^+]_S\vert (\bar u b)_{V-A}|\bar B^0\right\rangle \cdot \left\langle K^-|(\bar s u)_{V-A}|0 \right\rangle,
\ee
where the short distance part, similar to  Eq.~(6) of Ref.~\cite{PLB699_102}, is
\be \label{RSpi}
\bar y=\lambda_u^{(s)}\left\{a_{1y}
  +  a_{4y}^u  +a_{10y}^u-\left[ a_{6y}^u +a_{8y}^u\right]  r_\chi^K\right\}
 + \lambda_c^{(s)} \left\{
  a_{4y}^c  +a_{10y}^c-\left[ a_{6y}^c +a_{8y}^c\right]  r_\chi^K\right\}.
\ee
In the evaluation of the long distance matrix element $ \left\langle [\bar K^0 K^+]_S\vert(\bar u b)_{V-A}|\bar B^0\right\rangle$,  we assume that the transitions of $\bar B^0$ to the  $[\bar K^0K^+]_{S}$ states go first through intermediate meson resonances  $R_{S}$  which then decay into a $\bar K^0K^+$ pair.
This decay is described by a vertex function $G_{R_S[\bar K^0K^+]}(s_+)$.
For the intermediate resonances, as can be seen in Table~\ref{Tabfsi},
we have $R_S\equiv a_0(980)^+$ and $a_0(1450)^+$. Then using Eqs.~(B1) and~(B6) of Ref.~\cite{ChengPRD76_094006} Eq.~(\ref{M1L1}) leads to
 \begin{eqnarray}
 \label{M1S1}
\bar{A}_2(s_{0},s_{-},s_{+})
& =& - \frac{G_F}{2}\ f_K  \ (m_{B^0}^2- s_+) \ \sum_{R_S}F_0^{\bar B^0 R_S[\bar K^0K^+]}(m_{K}^2) 
\ \bar y(R_S K^-)\nonumber \\
&\times& G_{R_S[\bar K^0K^+]}(s_+)\  \langle R_S[\bar K^0K^+] \vert u \bar{d}\ \rangle,
\end{eqnarray}
$f_K$ being the charged kaon decay constant (Table~\ref{fixed}). 
Assuming that the variation of the $\bar B^0$ to $R_S$ transition form factor from one resonance to the other is small, we choose $R_S$ to be $a_0(980)^+$ which we denote  as $a_0^+$.
We can then parametrize the sum over the $R_S$ resonances by\footnote{This parametrization is quite similar to that of Eq.~(20) introduced in Ref.~\cite{JPD_PRD103} for the $D^0$ case.}

\begin{eqnarray}\label{sumRS}
 &&\sum_{R_S}F_0^{\bar B^0 R_S[\bar K^0K^+]}(m_{K}^2) \ 
\bar y(R_S K^-) \ G_{R_S[\bar K^0K^+]}(s_+)\  \langle R_S[\bar K^0K^+] \vert u \bar{d}\rangle  \nonumber \\
&& \simeq F_0^{\bar B^0 a_0^+}(m_{K}^2)\  \bar y(a_0^+ K^-)\ G_1(s_+) 
\end{eqnarray}
where we use
\be\label{RSa0+}
\langle R_S[\bar K^0K^+] \vert u\bar{d}\ \rangle=\langle a_0^+\vert u\bar{d}\ \rangle=1.
\ee
\begin{figure}[h]  \begin{center}
\includegraphics[scale=0.65]{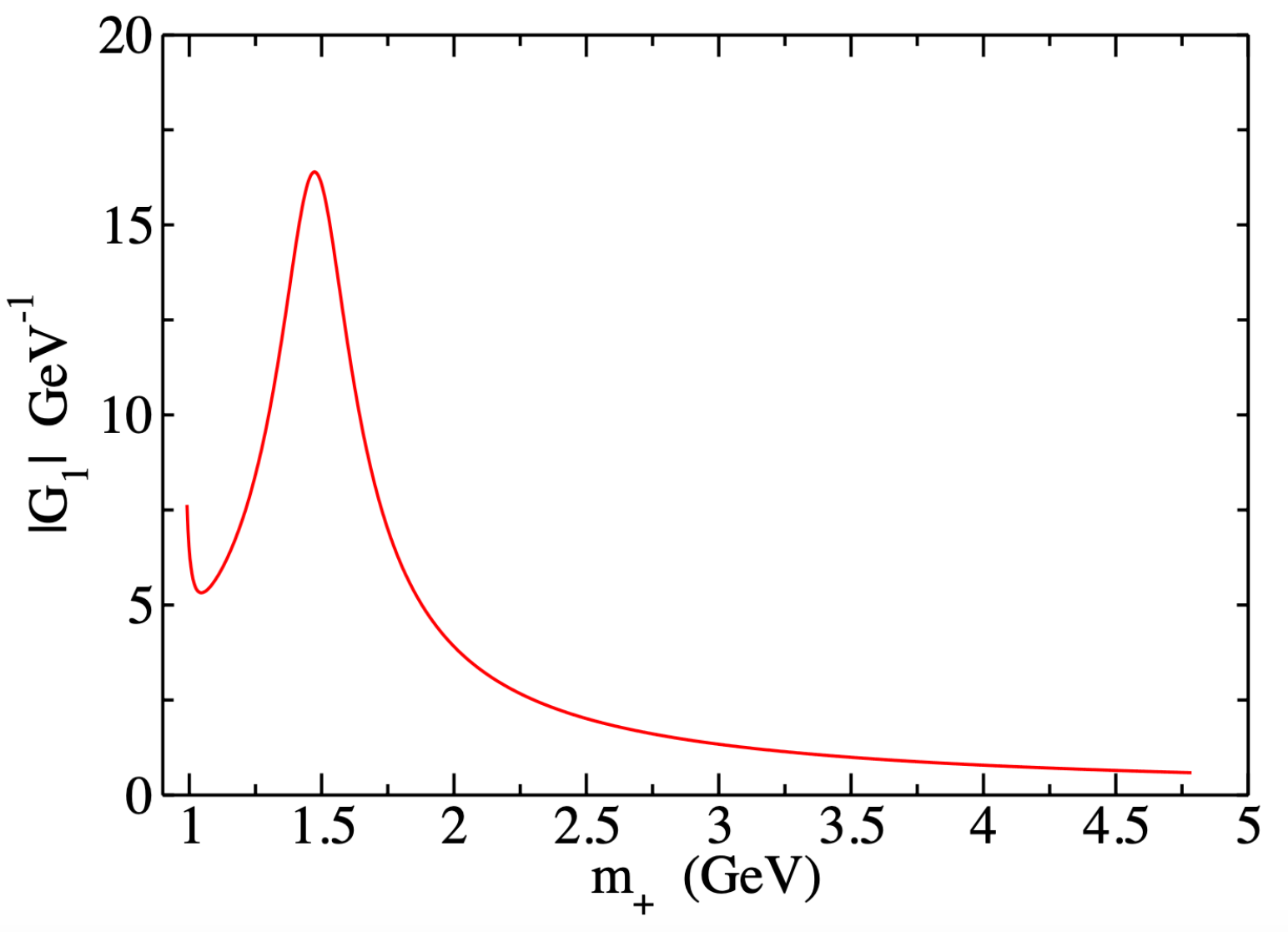}
\caption{Modulus of the $|G_1(s_+)|$ ($m_+=\sqrt{s_+}$) function  which describes the $\bar B^0$ transition to the scalar-isovector $\bar K^0 K^+$ state.
The threshold enhancement is due to the $a_0(980)^+$ resonance and the peak around~1.5~GeV to the $a_0(1450)^+$ one.
}
\label{ModG1}
\end{center} \end{figure}

The function $G_1(s)$ describes the  transition from a $u\bar d$ pair into a $\bar K^0K^+$ state.
It is  calculated from a unitary model with relativistic equations for the two-coupled channels $\pi \eta$ and $K \bar K$.
 It is based on the  two-channel model of the  $a_0(980)$ and $a_0(1450)$ resonances built in~Refs.~\cite{AFLL, AFLL2}. 
Details on its calculation are given in chapter IV of Ref.~\cite{JPD_PRD103}, in particular, see Eqs.~(104) to~(111).
The $G_1(s)$ function depends on two parameters $r_1$ and $r_2$ 
 which represent the coupling constants to the $\pi^+ \eta$ and $\bar K^0 K^+$ states, respectively.
In our model,  $r_2$ is taken as a free parameter with $r_1/r_2=0.88$ as in Ref.~\cite{JPD_PRD103}, keeping however the third-degree polynomial  $W(s)$ fixed to 1.
The modulus of the $G_1(s)$  function used in the present model is plotted in Fig.~\ref{ModG1}.
From Eqs.~(\ref{M1S1}) and~(\ref{sumRS}) we get the following contribution to the $\bar{B^0}\to K^0_S K^+ K^-$  amplitude\footnote{ 
An alternative~\cite{PRD96_113003} could be to parametrize the $\bar B^0$ transition to $\bar K^0 K^+$ as being proportional to  the scalar-isovector form factor.
This form factor has been calculated in Ref.~\cite{Bachir2015} using  MO dispersion relation approach~\cite{Barton1965, MO}.}
\be\label{A2b}
\bar{A}_2(s_{0},s_{-},s_{+}) =  - \frac{G_F}{2}\ f_K \ \bar{y}(a_0^+ K^-) \  (m_{B^0}^2- s_+)\  F_0^{\bar B^0 a_0^+}(m_{K}^2)\ G_1(s_+),
\ee
Charge conjugation transformation applied to Eq.~(\ref{A2b}) gives the following contribution for the $B^0$ case,
\be\label{A2}
{A}_2(s_{0},s_{+},s_{-}) =  - \frac{G_F}{2} \ f_K \ {y}(a_0^- K^+) \  (m_{B^0}^2- s_-)\  F_0^{B^0 a_0^-}(m_{K}^2)\ G_1(s_-),
\ee
 where $ {y}=\bar{y}$ with
$\lambda_p^{(s)} \to \lambda_p^{(s)* }|_{p=u,c} $ and $ F_0^{B^0 a_0^-}(m_{K}^2)=F_0^{\bar B^0 a_0^+}(m_{K}^2)$.

\subsection{Contributions to the amplitude with two kaons in $P$ wave}
\label{2kaonPwave}
\subsubsection{The $K^+K^-$ contributions}

Retaining the part coming  from ${\cal{H}}^{(p)}_3$  [see Figs.~1(b) and~2(a)] one has for this term of the $\bar{B^0}\to~K^0_S~K^+K^-$ amplitude [Eqs.~(\ref{Mfull}) and (\ref{TpK0KK})],
\be \label{M3LI}
{\bar {\cal{A}}}_{3L,I}(s_{0},s_{-},s_{+}) =  \frac{G_F}{2}\ \bar w_u(\bar K^0 R_L^I)\ \langle \bar K^0|(\bar s b)_{V-A}|\bar B ^0\rangle \cdot \langle [K^+ K^-]_L^I |(\bar u u)_{V-A}|0\rangle,
\ee
with (see also Eq.~(8) in Ref.~\cite{PLB699_102})
\be \label{wu}
\bar w_u = \lambda_u^{(s)}\ a_{2w} + 
\left ( \lambda_u^{(s)} +  \lambda_u^{(s)} \right )\ \left (a_{3w} + a_{5w} +a_{7w} + a_{9w} \right).
\ee
and (Eq.~(5) in Ref.~\cite{PLB699_102})
\be \label{Fu}
\langle [K^+(p_+) K^-(p_-)]_L^I |(\bar u u)_{V-A}|0\rangle = (p_+ - p_-)\ F_u^{[K^+K^-]_L^I}.
\ee
In the above term only $P$-waves contribute. Following Eq.~(B6) in Ref.~\cite{ChengPRD76_094006} for the evaluation of the matrix element $\langle \bar K^0|(\bar s b)_{V-A}|\bar B ^0\rangle $, we obtain 
\be \label{M3PI}
{\bar {\cal{A}}}_{3P,I}(s_{0},s_{-},s_{+}) =  \frac{G_F}{2}\ (s_+ - s_-)\ \bar w_u(\bar K^0 R_P^I)\ F_1^{\bar B^0 \bar K^0}(s_0) \ F_u^{[K^+K^-]^I}(s_0)
\ee
with $I=0$ or $1$.
For the  vector $\bar B^0 \bar K^0$ transition form factor,  one can use, as in Ref.~\cite{PLB699_102}, the parametrization given by Eq.~(30) of Ref.~\cite{Ball_PRD71_014015},
\be \label{F1barB0}
 F_1^{\bar B^0 \bar K^0}(s_0) = \frac{r_1}{1-\frac{s_0}{m_1^2}} + \frac{r_2}{(1-\frac{s_0}{m_1^2})^2} 
\ee
with $r_1= 0.162, r_2=0.173$ and $m_1=5.41$ GeV.

Reference~\cite{Bruch_2005} provides an evaluation of the form factor $F_u^{[K^+K^-]^I}(s_0)$ using vector dominance, quark model assumptions and isospin symmetry. 
It receives contributions from the $I=0$, {$\omega(782), \ \omega(1420)$ and $\omega(1650)$} resonances as well as those from  the $I=1$, $\rho(770), \rho(1450)$ and $\rho(1700)$ resonances\footnote{
 In the following the several $\omega$ and $\rho$ resonances wiil be denoted as $\omega, \omega{'}, \omega{''}$
 and $\rho, \rho{'}, \rho{''}$, respectively.}.
  Following Eq.~(23) of Ref.~\cite{PLB699_102},
\be \label{FuK+K-I}
F_u^{K^+K^-}(s_0) = F_u^{[K^+K^-]^{I=0}}(s_0) +F_u^{[K^+K^-]^{I=1}}(s_0),
\ee
with
\be \label{FuK+K-I0}
F_u^{[K^+K^-]^{I=0}}(s_0) =\frac{1}{2}  
\left [c_\omega^K BW_\omega(s_0) + c_{\omega{'}}^K BW_{\omega{'}}(s_0) +  c_{\omega{''}}^K BW_{\omega{''}}(s_0) \right ],
\ee
and
\be \label{FuK+K-I1}
F_u^{[K^+K^-]^{I=1}}(s_0) =\frac{1}{2}  
\left [c_\rho^K BW_\rho(s_0) + c_{\rho{'}}^K BW_{\rho{'}}(s_0) +  c_{\rho{''}}^K BW_{\rho{''}}(s_0) \right ].
\ee
Here the $BW_{R_L^I}(s_0)$ are the energy-dependent Breit-Wigner functions defined for each resonance $R_L^I$ of mass $m_{R_L^I}$ and width $\Gamma_{R_L^I}$ as
\be\label{BW}
BW_{R_L^I}(s_0)= \frac{m_{R_L^I}^2}{m_{R_L^I}^2-s_0-i \sqrt{s_0} \  \Gamma_{R_L^I}}.
\ee
The $c^K_{R_L^I}$  parameters have been determined in Ref.~\cite{Bruch_2005} through a constrained fit to the electromagnetic kaon form factors and we use the values given in their Table~2.
 
The fifth term, ${\cal{H}}^{(p)}_5$ [see Fig.~2(b)], in Eqs.~(\ref{TpK0KK}) yields also only a $P$-wave contribution,
\be \label{M5PI}
{\bar {\cal{A}}}_{5P,0}(s_{0},s_{-},s_{+}) =  \frac{G_F}{2}\ (s_+ - s_-)\  \bar w_s(\bar K^0 R_P^0)  \  F_1^{\bar B^0 \bar K^0}(s_0) \ F_s^{K^+K^-}(s_0),
\ee
with (see also Eqs.(~10) in Ref~\cite{PLB699_102})
\be\label{ws}
\bar w_s= \left ( \lambda_u^{(s)} +  \lambda_u^{(c)} \right )
 \left [a_{3w}+ a_{5w}- \frac{1}{2}\ \left (a_{7w}+ a_{9w}\right ) \right ]
+ \lambda_u^{(s)} \left (a_{4w}^u - \frac{1}{2} a_{10w}^u\right )+
  \lambda_c^{(s)} \left (a_{4w}^c - \frac{1}{2} a_{10w}^c\right).
\ee
The form factor $ F_s^{K^+K^-}(s_0)$,  described in terms of the $\phi(1020)$ and $\phi(1680)$ resonances denoted as $\phi$ and $\phi'$,  is given by 
(see Ref.~\cite{Bruch_2005} and also  Eq.~(25) of Ref.\cite{PLB699_102})
\be \label{FsK+K-0}
F_s^{K^+K^-}(s_0) = - c_{\phi}\  BW_{\phi}(s_0) - c_{\phi '}\ BW_{\phi '}(s_0).
\ee
As above for the contributions of the $\omega$ and $\rho$ resonances, the $\phi$ Breit-Wigner functions are given by Eq.~(\ref{BW}) and the $c_{\phi(\phi')}$ coefficients by the constrained fit results of Table~2 of Ref.~\cite{Bruch_2005}.

Adding the contributions of Eqs.~(\ref{M3PI}) and~(\ref{M5PI}) gives for the $\bar B^0$ case,
\bqa
\label{M35P0b}
\bar{A}_3(s_{0},s_{-},s_{+}) &\equiv& \sum_{I=0,1}{\bar {\cal{A}}}_{3P,I} (s_0,s_-, s_+)
+ {\bar {\cal{A}}}_{5P,0} (s_0,s_-, s_+)\nonumber \\
&=& -  \frac{G_F}{2}(s_- - s_+) F_1^{\bar B^0 \bar K^0}(s_0)  
 \left (\bar{w}_uF_u^{K^+K^-}(s_0)
+ \bar{w}_s F_s^{K^+K^-}(s_0) \right).
\eqa
The corresponding  $B^0$ part  is
\be
\label{M35P0}
{A}_3(s_{0},s_{+},s_{-})
=  - \frac{G_F}{2}(s_- - s_+) F_1^{ B^0 K^0}(s_0) \left ({w}_u F_u^{K^-K^+}(s_0) +{w}_s F_s^{K^-K^+}(s_0) \right),
\ee

with $F_1^{B^0 K^0}(s_0) = - F_1^{\bar B^0 \bar K^0}(s_0)$,  $w_{u,s}=\bar w_{u,s}(\lambda_p^{(s)} \to \lambda_p^{(s)*}|_{p=u,c} )$, $F_{u(s)}^{K^-K^+}(s_0)=F_{u(s)}^{K^+K^-}(s_0)$.

\subsubsection{The $K^0_S K^\pm$ contributions}
From the  ${\cal{H}}^{(p)}_1$ term, using Eq.~(B6) of Ref.~\cite{ChengPRD76_094006} together with relations similar to those of the Eqs.~(A.15) to (A.19) of Ref.~\cite{DedonderPol}, one obtains, 
 for the vector-isovector $[\bar{K}^0 K^+] _{P} \ K^- $  mode, the following contribution to the $\bar B^0$ amplitude (see Figs.~1(a) and 3)
\bqa \label{M1P1}
 \bar{A}_4 (s_{0},s_{-},s_{+})\equiv {\bar {\cal{A}}}_{1P,1}(s_{0},s_{-},s_{+})&=& - \frac{G_F}{2}\ f_K \ \left (s_0 - s_- + (m_{B^0}^2- m_{K}^2)\ \frac{m_{K^0}^2-m_{K}^2}{s_+} \right ) \nonumber \\
&\times& \sum_{R_P} A_0^{\bar B^0R_P[\bar{K}^0 K^+]}(m_{K}^2) \ m_{R_P[\bar{K}^0 K^+]}\ \bar y(R_P K^-) \nonumber \\\
 &\times&G_{R_P[\bar{K}^0 K^+] }(s_+) \ \langle R_P[\bar{K}^0 K^+]\vert u \bar{d}\rangle
\eqa
where $\langle R_P[\bar{K}^0 K^+] \vert u \bar{d}\rangle= 1$ since it is associated to 
 the $\rho(770)^+$, $\rho(1450)^+$ and $\rho(1700)^+$ resonances.
The sum over the vertex functions $G_{R_P[\bar{K}^0 K^+] }(s_+) $ can be parametrized using the  vector-isovector form factor~\cite{PRD96_113003}  $ F_1^{\bar{K}^0 K^+}(s_+) $ and,
\bqa \label{sumR1P}
 \sum_{R_P} \hspace{-0.3cm} &&A_0^{\bar B^0K^-}(s_+) \ m_{R_P[\bar{K}^0 K^+]}\ \bar y(R_PK^-) \ G_{R_P[\bar{K}^0 K^+] }(s_+)\ \langle R_P[\bar{K}^0 K^+] \vert u \bar{d}\rangle \nonumber \\ &&= \frac {\bar y(\rho K^-)}{f_\rho} A_0^{\bar B^0 \rho^+}(m_{K}^2)\ F_1^{\bar{K}^0 K^+}(s_+), 
\eqa
with the choice $\rho \equiv \rho(770)$ and $f_\rho$ being the charged $\rho$ decay constant  (Table~\ref{fixed}).
From Eqs.~(\ref{M1P1}) and (\ref{sumR1P}) one gets for the $\bar B^0$
 \begin{eqnarray} 
 \label{M1Pyb}
 \bar{A}_4 (s_{0},s_{-},s_{+})
&=& - \frac{G_F}{2}\ \frac{f_{K}}{f_{\rho} }\ \left (s_0 - s_- + (m_{B^0}^2- m_{K}^2)\ \frac{m_{K^0}^2-m_{K}^2}{s_+} \right ) \nonumber\\
& \times& \bar{y}(\rho^+ K^-) \ A_0^{\bar B^0 \rho^+}(m_{K}^2) F_1^{\bar{K}^0 K^+}(s_+),
 \end{eqnarray} 
The Wilson coefficient combination $\bar{y}(\rho^+ K^-)$ is given by Eq.~(\ref{RSpi}).
The value used for the $A_0^{\bar B^0 \rho^+}(m_{K}^2)$ transition form factor, determined in Ref.~\cite{bene03}, is given in Table~\ref{fixed}.
As shown in Ref.~\cite{Bruch_2005} the form factor, $F_1^{\bar{K}^0 K^+}(s_+)=2~F_u^{[K^+K^-]^{I=1}}(s_+)$ gets contributions from the three $\rho$ resonances [see Eq.~(\ref{FuK+K-I1})].

The $B^0$ part reads
 \begin{eqnarray} 
 \label{M1Py}
 {A}_4 (s_{0},s_{+},s_{-})
&=& - \frac{G_F}{2}\ \frac{f_{K}}{f_{\rho}} \ \left [s_0 - s_+ + (m_{B^0}^2- m_{K}^2)\ \frac{m_{K^0}^2-m_{K}^2}{s_-} \right ] \nonumber\\
& \times& {y}(\rho^- K^+) \ A_0^{B^0 \rho^-}(m_{K}^2) F_1^{{K}^0 K^-}(s_-),
 \end{eqnarray} 
with ${y}(\rho^- K^+)=\bar {y}(\rho^+ K^-); \lambda_p^{(s)}|_{p=u,c} \to \lambda_p^{(s)*}|_{p=u,c}$,  $A_0^{B^0 \rho^-}(m_{K}^2) =-A_0^{\bar B^0 \rho^+}(m_{K}^2)$ and $F_1^{{K}^0 K^-}(s)=-F_1^{\bar {K}^0 K^+}(s)$.

\subsection{Contributions to the amplitude with $K^0_S K^\pm$ states in $D$ wave}
\label{DwaveAmplitude}
One cannot form a two-kaon $D$-wave state  from the vacuum state through the $(\bar q q)_{V-A}$ operator, consequently there is no such part arising from the ${\cal{H}}^{(p)}_i$ terms for $i=3, 5$ and  6.
Here the contribution coming from the  ${\cal{H}}^{(p)}_1$ term  (see Figs.~1(a) and~3) with  a two-kaon $D$-wave state, saturated by the $a_2(1320)^+ $ resonance, reads (see e.g. Eq.~(A.23) of Ref.~\cite{DedonderPol}),
\bqa
\bar{A}_5 (s_{0},s_{-},s_{+})\equiv{\bar {\cal{A}}}_{1D,1} (s_0, s_-,s_+) & = & - \frac{G_F}{2} \ f_K \ {\bar{D}(\bf{p_0}, \bf{p_-})}  \sum_{R_D\equiv a_2^+} 
F^{\bar B^0 R_D [\bar K^0 K^+]}(m_K^2, s_+) \nonumber \\
&\times& \bar y(R_D K^-) \ G_{R_D [\bar K^0 K^+]}(s_+)\ \langle R_D [\bar K^0 K^+] \vert u\bar d \rangle.
\eqa
With $\langle a_2^+ [\bar B^0 K^+] \vert u\bar d \rangle=1$ one obtains  for the $\bar B^0$ case
\be \label{M1D1tp}
\bar{A}_5 (s_{0},s_{-},s_{+}) = - \frac{G_F}{2} \ f_K \ \bar{y}(a_2^+ K^-) \ g_{a_2^+\bar K^0 K^+} \ 
\frac{{\bar D(\bf{p_0}, \bf{p_-})}}{m_{a_2}^2 - s_+ - i\ m_{a_2} \ \Gamma_{a_2}(s_+)} \ F^{\bar B^0 a_2^+}(m_K^2, s_+),
\ee
where the Wilson coefficient combination $\bar{y}(a_2^+ K^-)$ is given by Eq.~(\ref{RSpi}).
The coupling constant $g_{a_2^+\bar K^0 K^+}$ characterizes the strength of the $a_2^+\to~\bar K^0 K^+$ transition.
 The function $\bar D(\bf{p_0}, \bf{p_-})$ is defined by
\be \label{D0-}
 {\bar D(\bf{p_0}, \bf{p_-})} = \frac{1}{3} \ (\vert \bf{p_0} \vert \vert \bf{p_-} \vert)^2 - (\bf{p_0} \cdot \bf{p_-})^2.
\ee
In the $\bar K^0 K^+$ center-of-mass system the moduli of the $\bar{K}^0$ and $K^-$ momenta are given by
\bqa \label{modp1-}
{\vert {\bf{p_0}} \vert} & = & \frac{1}{2} \ \sqrt{ \frac{[s_+ - ( m_K + m_{K^0})^2] \ [s_+ -( m_K - m_{K^0})^2]}{s_+}},\nonumber \\
{\vert {\bf{p_-}} \vert}  & = & \frac{1}{2} \ \sqrt{ \frac{[m_{B^0}^2- (\sqrt{s_+} + m_K)^2]\ [m_{B^0}^2- (\sqrt{s_+} - m_K)^2]}
{s_+}},
\eqa
and 
\be \label{p1dotp-}
4\ {\bf{p_0}} \cdot {\bf{p_-}} = s_0 - s_- + \frac{(m_{B^0}^2 -m_K^2) \ (m_{K^0}^2 - m_K^2)}{s_+}.
\ee
The transition form factor $F^{\bar B^0 a_2^+}(m_K^2, s)$ follows from Ref.~\cite{KimPRD67} and reads
\be\label{KimD}
 F^{\bar B^0 a_2^+}(m_K^2, s) = k^{\bar B^0 a_2^+}(m_K^2) + b_+^{\bar B^0 a_2^+}(m_K^2) \ (m_{B^0}^2 - s) 
 +   b_- ^{\bar B^0 a_2^+}(m_K^2) \ m_K^2.
 \ee
 The form factors, $k^{\bar B^0 a_2^+}(m_K^2)$ and $b_{\pm}^{\bar B^0 a_2^+}(m_K^2)$  are not known.
 In our model we will fix $s$ in Eq.~(\ref{KimD}) to the $a_2$ resonance mass squared and the value we use is given in Table~\ref{fixed}. For the $B^0$ case, we have
\be \label{M1D1t}
{A}_5 (s_{0},s_{+},s_{-})
  = - \frac{G_F}{2} \ f_K \ {y}(a_2^-  K^+) \ g_{a_2^- K^0 K^-} \ 
\frac{{D(\bf{p_0}, \bf{p_+})}}{m_{a_2}^2 - s_- - i\ m_{a_2} \ \Gamma_{a_2}(s_-)} \ F^{ B^0 a_2^-}(m_K^2, m_{a_2}^2),
\ee
with ${y}(a_2^-  K^+) =\bar{y}(a_2^+ K^-; \lambda_p^{(s)} \to \lambda_p^{(s)*}|_{p=u,c}) $, $g_{a_2^- K^0 K^-}=g_{a_2^+ \bar K^0 K^+}$ and $F^{ B^0 a_2^-}(m_K^2, m_{a_2}^2)=F^{ \bar B^0 a_2^+}(m_K^2, m_{a_2}^2)$.
The function $D(\bf{p_0}, \bf{p_+})$ of the $K^0$ and $K^+$  momenta in $K^0 K^+$ center-of-mass system is defined in a similar way to that of the function $\bar{D}(\bf{p_0}, \bf{p_-})$ in Eq.~(\ref{D0-}) but the variables $s_+$ and $s_-$ have to be interchanged.

\section{Results and discussion} \label{results}

The Belle~\cite{PRD82_073011} and \textit{BABAR}~\cite{PRD85_112010} Collaboration analyses of the  $B^0 \to K^0_S K^+ K^-$ data have been performed within a time-dependent-Dalitz approach.
As shown in Appendix~\ref{B0B0bmixingtimedep} [see Eq.~(\ref{d2BrbarA})] the double differential branching fraction or the Dalitz plot density distribution for the $\bar{ B^0}\to K^0_S K^+ K^-$ decay  can be written as \be \label{d2Brbar}
\frac{d^2{\rm Br(\bar{ B^0})}}{ds_+ds_0}=\frac{1}{32(2\pi)^3m_{B^0}^3 \Gamma_{B^0}}
[(1-x)|\bar{A}(s_0,s_-,s_+)|^2+x |A(s_0,s_+,s_-)|^2],
\ee
where $\bar{A}(s_0,s_-,s_+)=\sum_{i=1}^5{\bar{A}}_i(s_0,s_-,s_+)$ is our decay amplitude for the $\bar{ B^0}\to K^0_S K^+ K^-$ process, $A(s_0,s_+,s_-)=\sum_{i=1}^5A_i(s_0,s_+,s_-)$ is that for the $B^0$ decay and $\Gamma_{B^0}$ is the $B^0$ width. 
The different parts, $\bar{A}_i(s_0,s_-,s_+)$ and $A_i(s_0,s_+,s_-)$, of our decay amplitudes have been given in Sec.~\ref{SelectedAmplitudes}.
The parameter $x$ gives the strength of the contribution of the $B^0$-$\bar{B^0}$ transition process. 
It is equal to
\be \label{x}
x= \frac{1}{2}~\Big[~1-\frac{1-2 w}{(\Delta m_d/\Gamma_{B^0})^2+1}~\Big],
\ee
where $\Delta m_d$ is the difference of the heavy and light $B^0$ mass eigenvalues
and $w$ is the fraction of events in which the other $B^0$ meson is tagged with the incorrect flavor~\cite{PRD85_112010}.
The double differential branching fraction or the Dalitz plot density distribution for the $B^0\to K^0_S K^+ K^-$ decay reads
\be \label{d2Br}
\frac{d^2{\rm Br(B^0)}}{ds_+ds_0}=\frac{1}{32(2\pi)^3m_{B^0}^3 \Gamma_{B^0}}
[\,(1-x)|A(s_0,s_+,s_-)|^2+x |\bar{A}(s_0,s_-,s_+)|^2\,].
\ee
Here, compared to the $\bar B^0$ case, the amplitude arguments $s_-$ and $s_+$ are interchanged.

As in the Belle~\cite{PRD82_073011}  and \textit{BABAR}~\cite{PRD85_112010}  analyses the sum over both charge-conjugate-decay modes is implied, we compare the experimental effective 
$K^0_S K^+$, $K^0_S K^-$ and $K^+K^-$ mass projections with the corresponding theoretical distributions $dBr/ds_i$ obtained by a suitable integration over $s_0$ or $s_+$ of the sum of the differential branching fractions given by Eqs.~(\ref{d2Brbar}) and (\ref{d2Br}). 
Here  $s_i, i=1,2,3 $  denote the squares of the  three different $K^0_S K^+$, $K^0_S K^-$ and $K^+K^-$ effective masses of the final kaon pairs, respectively.

We have made a simultaneous fit of the model parameters to the Belle data presented in Fig.~3 of Ref.~\cite{PRD82_073011} and the \textit{BABAR} data shown in Fig.~17 of Ref.~\cite{PRD85_112010}.   
The background components have been subtracted to obtain the signal Belle distributions. 
We have also omitted the first data bins in the effective mass projections corresponding to the $s$ values smaller than their kinematical limits given by the masses of the $K \bar{K}$ pairs.
Among the Belle data, one has 76 points for the $K^0_S K^+$ mass distribution, 76 points for the $K^0_S K^-$ mass distribution, 149 points for the $K^+K^-$ mass distribution 
and 24 points concentrated in the narrow region of the $K^+K^-$ mass around the $\phi(1020)$ resonance. 
Each set of the three \textit{BABAR} distributions consists of 32 points. 
Altogether we have taken into account 325 Belle data points and 96 \textit{BABAR} data values.
As we fit also the branching fraction of the $B^0 \to K^0 K^+ K^-$ decay, the total number of the data points is equal to 422.

The theoretical values of the $K^0_S K^+$, $K^0_S K^-$ and $K^+K^-$ mass distributions $dN^{th}/dE_i$ have been related to the branching fraction distributions $dBr/ds_i$
using the relation
\be \label{Nth}
\frac{dN^{th}}{dE_i}= 2 E_i F_i \frac{dBr}{ds_i},
\ee
where $E_i=\sqrt s_i$ and
\be \label{norm}
F_i=\frac{N^{ev}_i d_i}{Br^{exp}}.
\ee
In this expression $N^{ev}_i$ is the total number of experimental events of a given distribution with the bin width $d_i$ and $Br^{exp}$ is the experimental branching fraction of the $B^0 \to K^0 K^+ K^-$ decay. For the description of the Belle data we use $N^{ev}_i=1125$ for every $i$ while for the \textit{BABAR} data sets we have $N^{ev}_1=1419$, $N^{ev}_2=1415$ and $N^{ev}_3=1449$ events.

\begin{table}
\caption{
Fitted strong interaction parameters of our model amplitude. 
The parameters  $C, P_i,\ i=3$~to~5  are dimensionless, $\phi_i, \ i=1,2$ are in radians, $ r_2$  in GeV$^{3/2}$ and the $c_i$, $i=1$~to 6 in GeV$^{-i}$.
 The component $\bar A_1$ is multiplied by the complex sixth order polynomial $P_1(z)= {\rm e}^{i \phi _1} C \left (1+\sum_{i=1}^{6} c_i~z^i \right )$ with $z=\sqrt{s_0}- 2 m_K$.
 The contribution $\bar A_2$, proportional to the function $G_1(s_+)$ where the parameter $r_2$ represents the coupling constant to the $\bar K^0 K^+$ state, is multiplied by the phase factor ${\rm e}^{i \phi _2}$. 
 For $j=3, 4, 5$ the real parameters $P_j$ renormalize the corresponding $\bar A_j$.
The same parameters are also introduced for the ${A_j}$, $j=1$~to~5, in the same way.
}
\begin{center}
\begin{tabular}{ccc}
\hline
\hline
Parameters & Values &\vspace*{0.2cm} \\ 
\hline
 $C$ & $ 0.84005$ &\vspace*{0.2cm} \\   
$\phi_1$ & $-3.4691$ rad & \vspace*{0.2cm} \\ 
$c_1$ & $-1.7509$ & \vspace*{0.1cm} \\ 
$c_2$ & $1.2298$ & \vspace*{0.1cm} \\ 
$c_3$ & $0.23169$ \vspace*{0.1cm} \\
$c_4$ & $-0.24359$ \vspace*{0.1cm} \\
$c_5$ & $0.064156$ \vspace*{0.1cm} \\
$c_6$ & $-0.0061211$ \vspace*{0.1cm} \\ 
$r_2$ & $8.6409$ GeV$^{3/2}$ & \vspace*{0.2cm} \\ 
$\phi_2$  & $ 4.4632$ rad & \vspace*{0.1cm} \\ 
$P_3$ & $1.1752 $ & \vspace*{0.2cm} \\  
$P_4$ & $0.38593$ & \vspace*{0.1cm} \\ 
$P_5$ & $0.29155$ & \vspace*{0.1cm} \\
\hline
\hline
\end{tabular}
\end{center}
\label{parameters}
\end{table}

 In our fit we use the $\chi^2$ function defined as
\be \label{chi2}
\chi^2= \sum_{j=1}^{421} 
\left[\frac
{\frac{dN^{th}}{dE}(E_j)-\frac{dN^{exp}}{dE}(E_j)}
{\Delta \frac{dN^{exp}}{dE}(E_j)}\right]^2
+\chi^2_{Br},
\ee
where 
\be \label{chiBr}
\chi^2_{Br}=w_{Br}\left[\frac{Br^{th}-Br^{exp}}{\Delta Br^{exp}}\right]^2,
\ee
$\frac{dN^{exp}}{dE}(E_j)$ is the experimental value of the mass distribution taken at $E_j$ and $\Delta\frac{dN^{exp}}{dE}(E_j)$ is its uncertainty while $\frac{dN^{th}}{dE}(E_j)$ is the corresponding theoretical value calculated at the same $E_j$.
We put $w_{Br}=20$ to get a good fit  for the theoretical $CP$-averaged branching fraction $Br^{th}$.

It turns out that to obtain a reasonable fit to the data one needs to modify the five components of the model amplitude.
The amplitudes $\bar  A_1$ and $A_1$ are multiplied by a sixth order polynomial  $P_1(z)$  of the variable  $z=\sqrt{s_0}- 2 m_K$  with
 \be \label{P1}
P_1(z)= {\rm e}^{i \phi _1} C \left (1+\sum_{i=1}^{6} c_i~z^i \right ).
\ee 
This introduces 8 real free parameters, $\phi_1$, $C$ and the $c_i, i=1$~to~6.
The  scalar-isovector $K^0 K^\pm$ terms $\bar  A_2$ and $A_2$ terms [Eqs. (\ref{A2b}) and~(\ref{A2})] are proportional to the $G_1(s)$ function in which the coupling constant $r_2$ [see the paragraph below Eq.~(\ref{RSa0+})] has been adjusted.
 Both terms have been multiplied by the phase factor $e^{i\phi_2}$, where $\phi_2$
is a real free parameter.
The $K^+ K^-$ and $K^0_S K^\pm$  $P$- wave components $\bar A_3$, $A_3$, $\bar A_4$,  $A_4$ and  the $K^0_S K^\pm$ $D$-wave $\bar A_5$ and $A_5$  ones need to be renormalized by the free real coefficients $P_3$, $P_4$ and $P_5$, respectively.

 \begin{figure}[h]  \begin{center}
\includegraphics[scale=0.29]{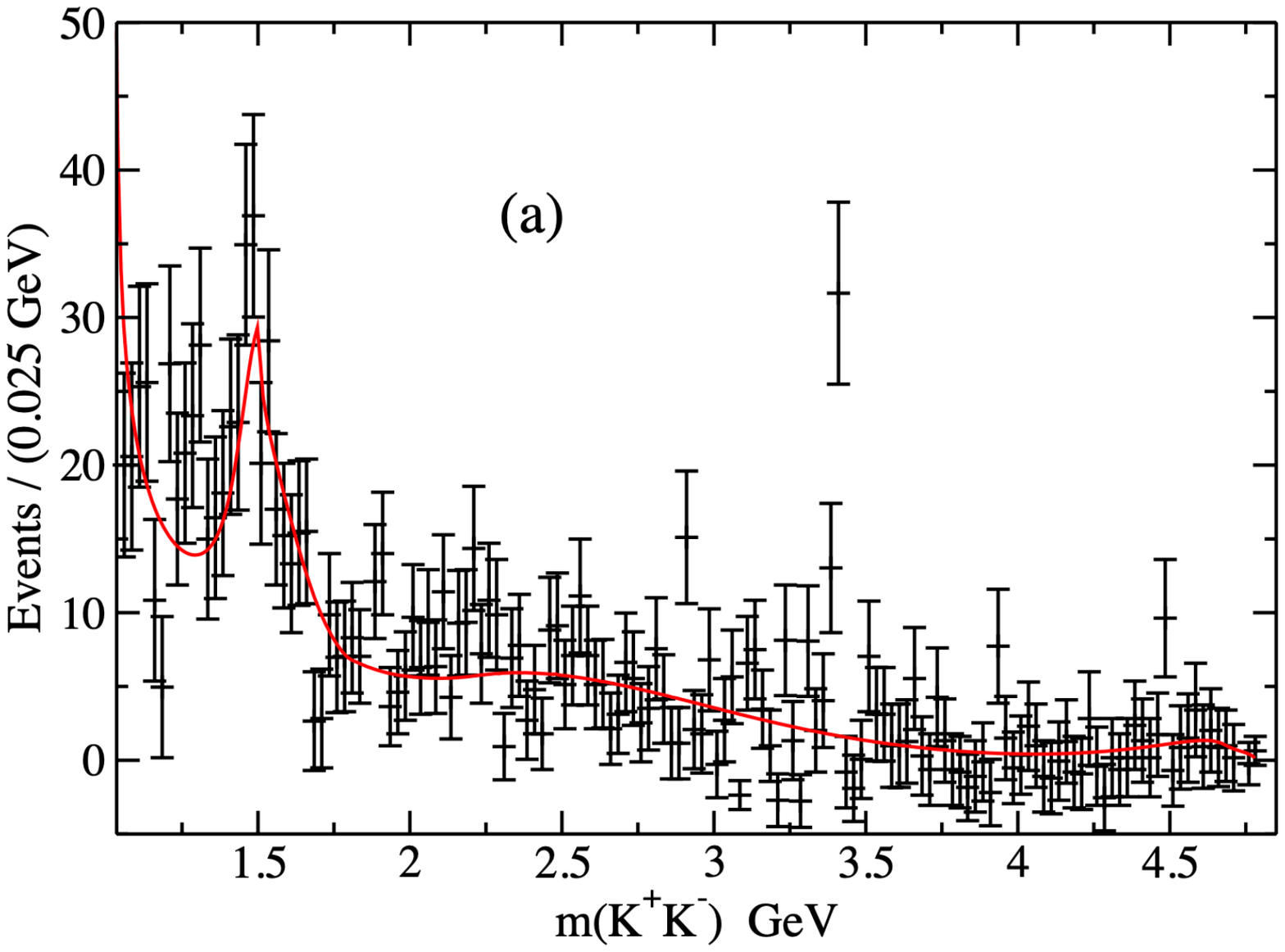}
\includegraphics[scale=0.433]{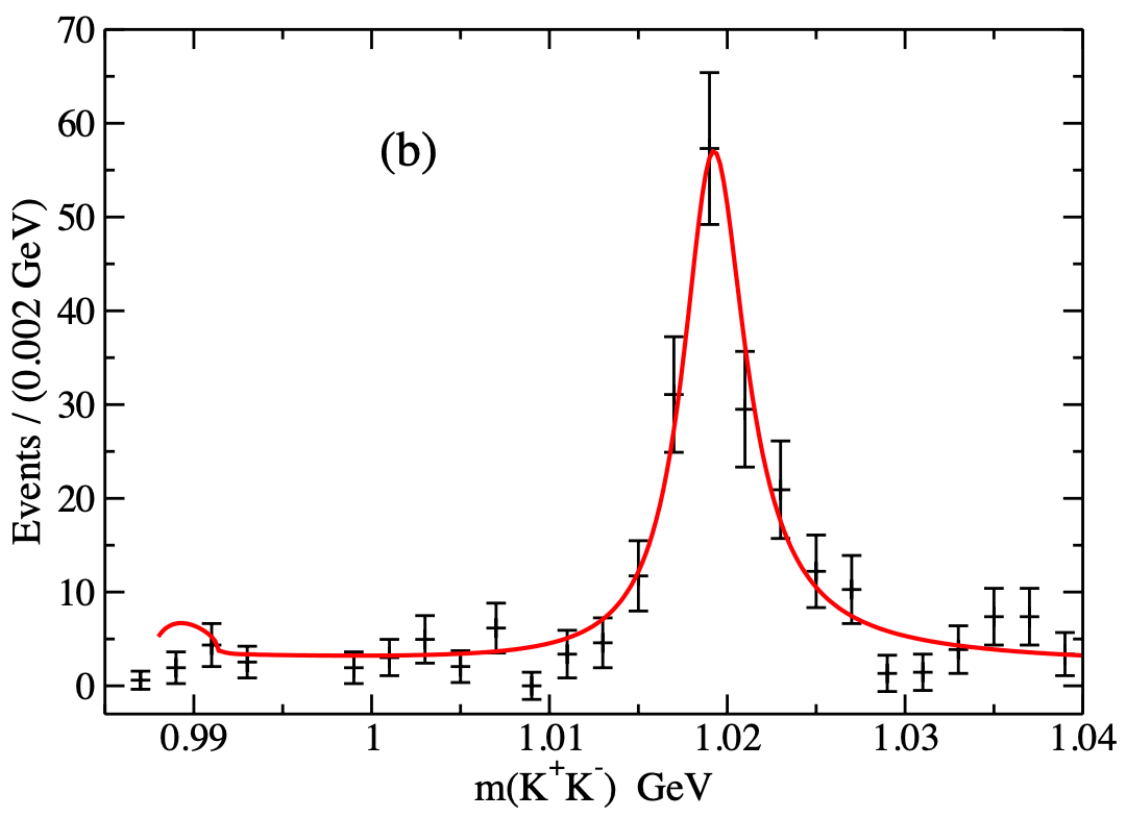}
\includegraphics[scale=0.433]{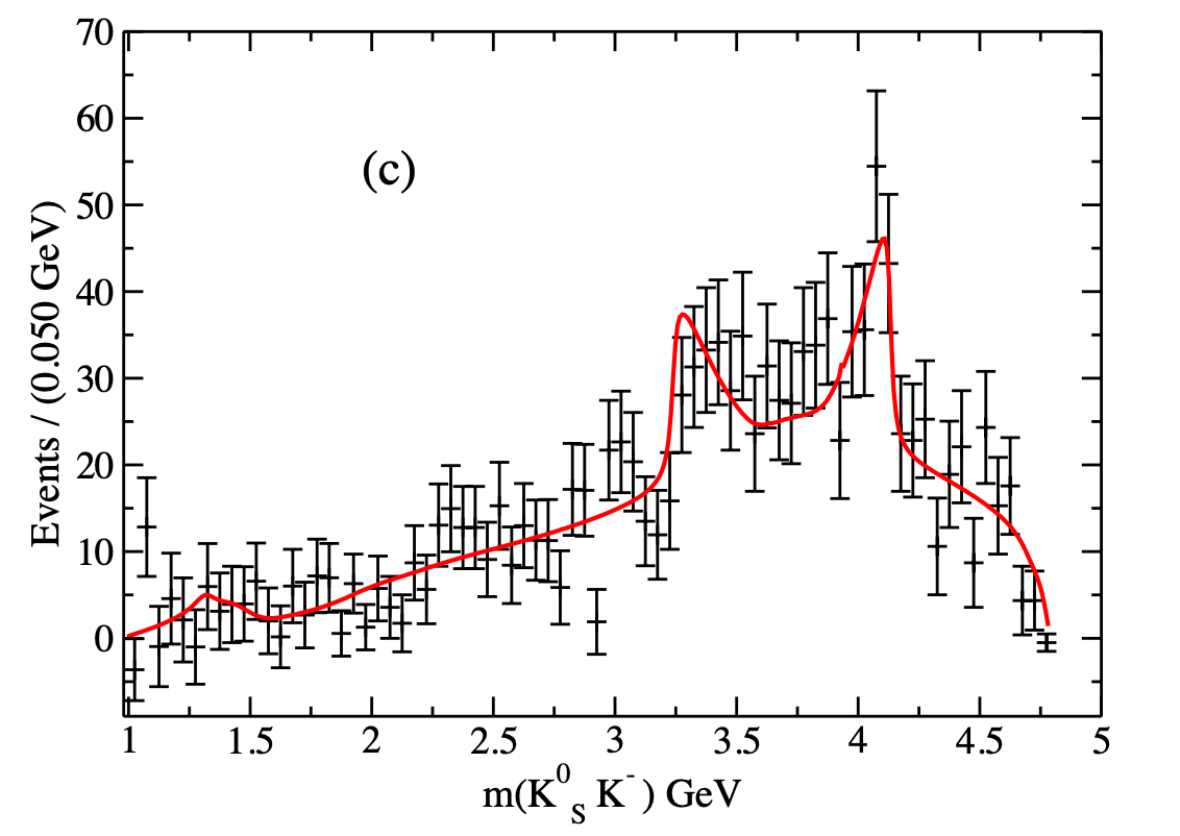}
\includegraphics[scale=0.433]{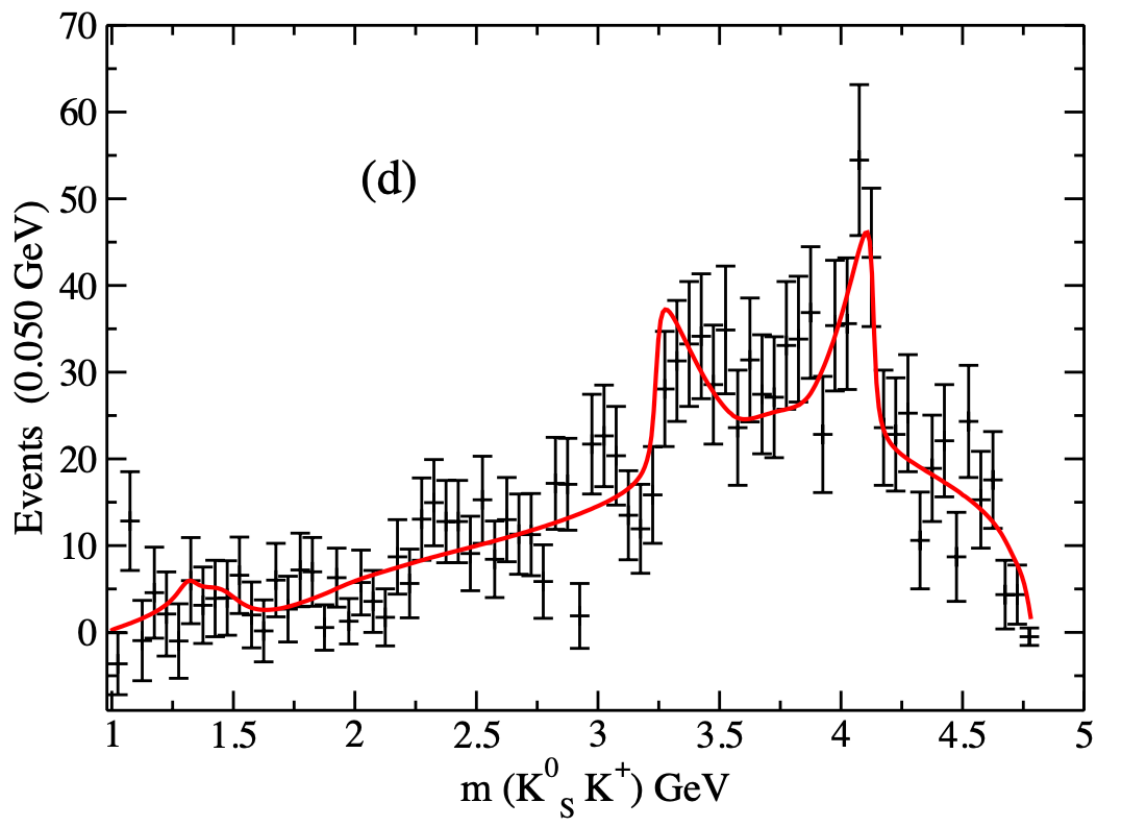}
\caption{The Dalitz-plot-projection fit (solid line) to the Belle~\cite{PRD82_073011} experimental data extracted from their Fig.~3: (a) the $m( K^+ K^-)$ distribution, (b) the $m( K^+ K^-)$ projection near the $\phi(1020)$ resonance, (c) the $m( K^0_S K^-)$ distribution and (d) that of the $m( K^0_S K^+)$.
The resonance $\chi_{c0}(1P)$ visible in the plot (a) at around 3.4 GeV has not been introduced in our amplitude.
The two bumps in (a) correspond to the $\phi(1020)$ and $f_0(1500)$, respectively; the bump in (b) close to the threshold comes from the $f_0(980)$; the first bump in (c)  and (d) is due to the $a_0(1450)$;  the two other ones in (c) and (d) are reflections of the $\phi(1020)$.
Data are represented by tiny horizontal lines with error bars. 
}\label{F5}
\end{center} \end{figure}
In  our fit, we use the measured ratio $\Delta m_d/\Gamma_{B^0}~=~0.769\pm0.004$~\cite{PDG2022} and we put the experimental parameter $w=0$, to get from Eq.~(\ref{x}) the value $x=0.186$. 
 The values of the 13 fitted parameters are given in Table~\ref{parameters}.
 We obtain $\chi^2=583.6$ which divided by the number of degrees of freedom, $ndf=409$, leads to $\chi^2/ndf=1.43$. 
The total $B^0\to K^0 K^+ K^-$ experimental branching fraction, $(2.68 \pm 0.11 )\times 10^{-5}$~\cite{PDG2022}, is very well reproduced as one gets the corresponding theoretical value equal to $Br^{th}=2.65\times10^{-5}$.

\begin{figure}[h]  \begin{center}
\includegraphics[scale=0.433]{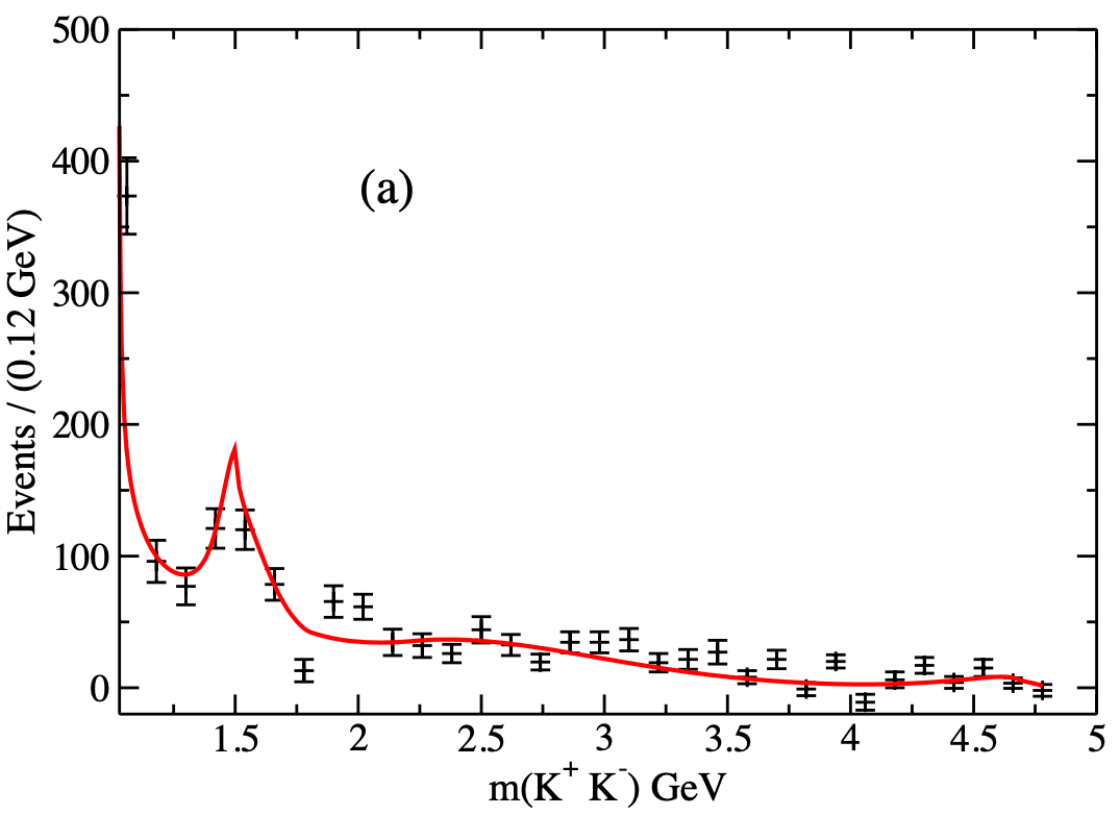}
\includegraphics[scale=0.433]{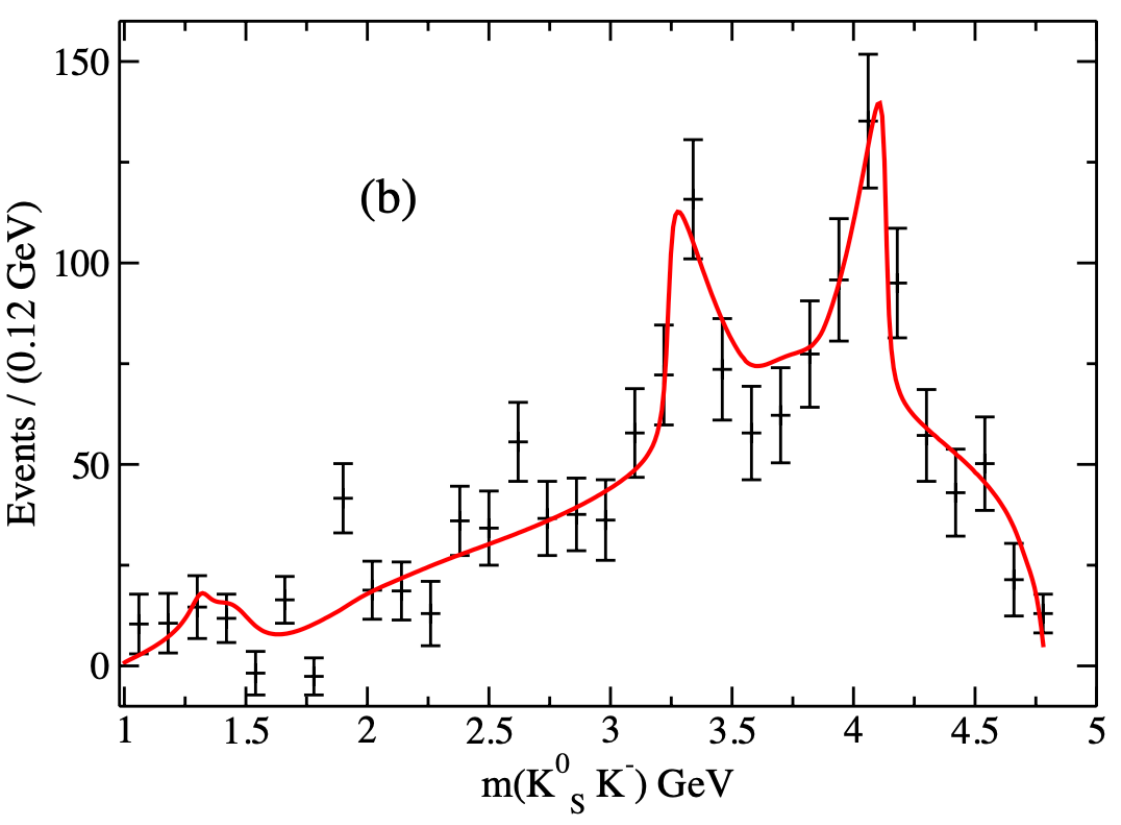}
\includegraphics[scale=0.433]{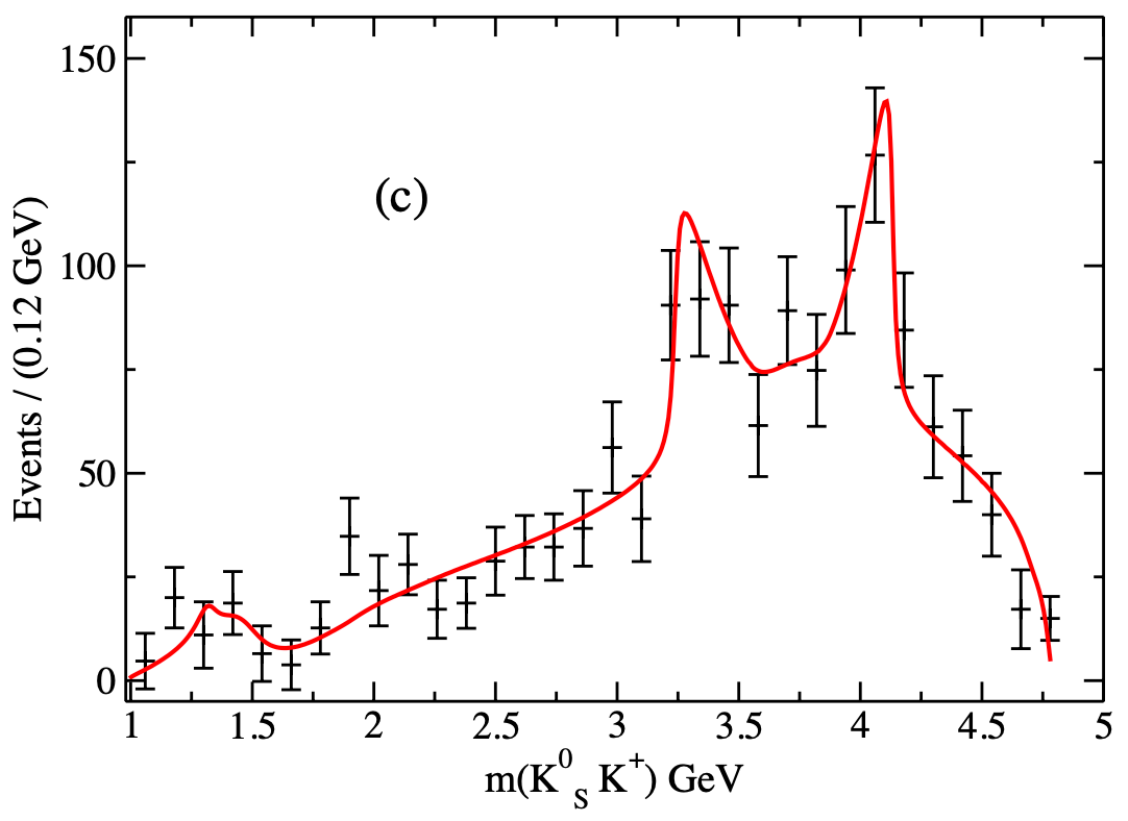}
\caption{  As in Fig.~\ref{F5} but for the \textit{BABAR}~\cite{PRD85_112010} data extracted from their Fig.~17: (a) the $m( K^+ K^-)$ distribution, (b) the $m( K^0_S K^-)$ one and (c) that of the $m( K^0_S K^+)$. The $\chi_{c0}(1P)$ signal is not visible in Fig.~17 of \textit{BABAR}.
  }\label{F6}
\end{center} \end{figure}
Our fit (solid line) to the mass projection distributions of the Belle~\cite{PRD82_073011}   and \textit{BABAR}~\cite{PRD85_112010}  experimental data is displayed in Figs.~\ref{F5} and~\ref{F6}, respectively.
The near threshold peak in the $m( K^+ K^-)$ distribution of the Belle Collaboration, Fig.~\ref{F5}~(a), is due to the $\phi(1020)$ and the next one to the $f_0(1500)$ denoted as $f_X$ in Ref.~\cite{PRD82_073011}.
The bump near 1.5~GeV in the plot (Fig.~\ref{ModGam2s}) of the modulus of the strange scalar form factor $\Gamma_2^s({s_0})$ contributes to this $f_X$ peak.
 Furthermore, in our model, it corresponds to the opening, close to  $2  m_\rho$, of the third effective $4 \pi$ channel~\cite{JPD_PRD103, EPJ}.
There are also some contributions from the $a_0(1450), \rho(1450)$ and $\omega(1420)$.
The resonance $\chi_{c0}(1P)$ visible at around 3.4 GeV in this Fig.~\ref{F5} (a)  has not been introduced in our amplitude.
In Fig.~\ref{F5} (b) the threshold bump arises from the $f_0(980)$ and the $\phi(1020$ peak is well reproduced.
In Figs.~\ref{F5}~(c) and~(d) the first bump comes from the $a_0(1450)$ and the two other ones are reflections of the $\phi(1020)$. 
Besides the fact that the projection distribution in the $\phi$ region has not been plotted and that the  $\chi_{c0}(1P)$  signal has not been kept, the \textit{BABAR} distributions, in Fig.~\ref{F6}, have characteristics similar to those of Belle.

 For the total branching fraction we obtain $Br^{th}(\bar{B^0}\to K^0_S K^+ K^-)=1.325\times10^{-5}$
which can be compared with $Br^{th}(B^0\to K^0_S K^+ K^-)=1.328\times10^{-5}$. The corresponding sum of these two branching fractions is equal to 
$Br=2.653\times 10^{-5}$.
Then the total {\it CP} asymmetry, 
\be\label{ACP}
A_{{CP}}=\frac{Br(\bar B^0)-Br(B^0)}{Br(\bar B^0)+Br(B^0)},
\ee
equals $- 0.11$~\%.
If one neglects the $B^0-\bar{B^0}$ transitions then this asymmetry becomes $A_{{CP}}=- 0.17$~\%.

The sum $Br_j$  of the integrated branching fractions for the $\bar{B^0}$  and $B^0$ decays into the $ K^0_S K^+ K^-$ system are calculated for the particular contributions of the modified $\bar A_j$ and $Aj$ terms.
The $Br_j$ values and the ratios $R_j=Br_j/Br$ are given in Table~\ref{rj} together with their sums  for $j=1$~to~5. 
We see that the $j=1$  term, with an  $S$-wave-$K^+K^-$ state, dominates with a contribution of~83.0~\% of the total branching fraction.
It arises  mainly from the $f_0(K^+K^-) K^0_S$ mode.
The second sizable contribution to $Br$, with~18.3~\% of the total, is the $j=3$ term with the $K^+K^-$ pair in $P$-wave.
It is dominated by $\phi K_S^0$ plus small $\omega K_S^0$ and $\rho^0 K_S^0$ modes.
Then follows the $a_0^\pm K^\mp$ mode  with~6.2~\%, the~$\rho^\pm K^\mp$ with~0.15~\% and the $a_2^\pm K^\mp$ with~0.11~\%.
The total percentage sum is~107.7~\% which indicates a small interference contribution.

\begin{table}
\caption{ Sum  $Br_j$   of the integrated branching fractions for $\bar{B^0}$  and  $B^0$ decays into the $ K^0_S K^+ K^-$ for the different modified $\bar {A}_j$ and ${A}_j$, their ratios $R_j$ (in \%) to the total branching fraction $Br=2.6516 \times10^{-5}$ and their sums for $j=1$~to~5.
}
\begin{center}
\begin{tabular}{ccccc} 
\hline
\hline
$j$  &        Final state modes    & Contributing & $Br_j$ &$R_j (\%)$\vspace*{-0.2cm} \\ 
       & $\bar B^0$ \ \ \ \ \ \ \ \ \ \ \ \ $B^0$   & resonances&             & \\        
\hline
$1$ & $[K^+ K^-]_S^0 \bar K^0 \ \ \ \ [K^- K^+]_S^0  K^0 $  &   $f_0$ & $2.20\times 10^{-5}$ &\  $83.0$\vspace*{0.2cm} \\   
$2$ & $[\bar K^0 K^+]_S^1 K^- \ \ \ \ [K^0 K^-]_S^1  K^+ $  &  $a_0^{\pm}$   & $1.64\times 10^{-6}$ &\ $6.2$\vspace*{0.2cm} \\ 
$3$ & $[K^+ K^-]_P^{0,1} \bar K^0 \ \ \ \ [K^- K^+]_P^{0,I}  K^0 $  & $\phi + \omega + \rho^0$    &\ $4.84\times 10^{-6}$ &$18.3$ \vspace*{0.2cm} \\ 
$4$ &  $[\bar K^0 K^+]_P^1 K^- \ \ \ \ [K^0 K^-]_P^1  K^+ $  &     $\rho^\pm$& $3.85\times 10^{-8}$ &$0.15$ \vspace*{0.2cm} \\ 
$5$ &   $[\bar K^0 K^+]_D^1 K^- \ \ \ \ [K^0 K^-]_D^1  K^+ $   &   $a_2^\pm$  & $2.87\times 10^{-8}$ &$0.11$ \vspace*{0.1cm} \\
\hline
\vspace*{0.1cm}
&  & &\hspace{-1. cm}$\sum_{j=1}^5\  2.86\times 10^{-5}$ &$107.7$  \vspace*{0.1cm} \\ 
\hline
\hline
\end{tabular}
\end{center}
\label{rj}
\end{table}
 \begin{figure}[h]  \begin{center}
\includegraphics[scale=0.43]{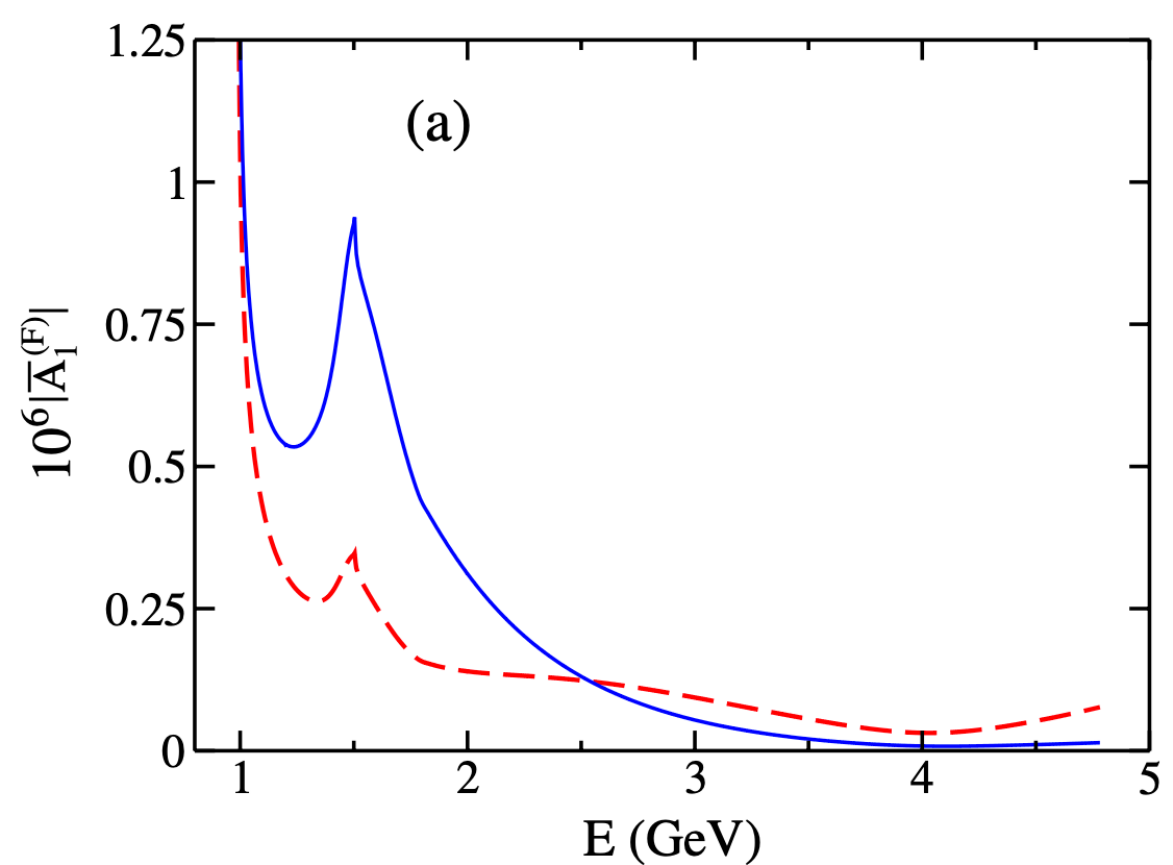}
\includegraphics[scale=0.43]{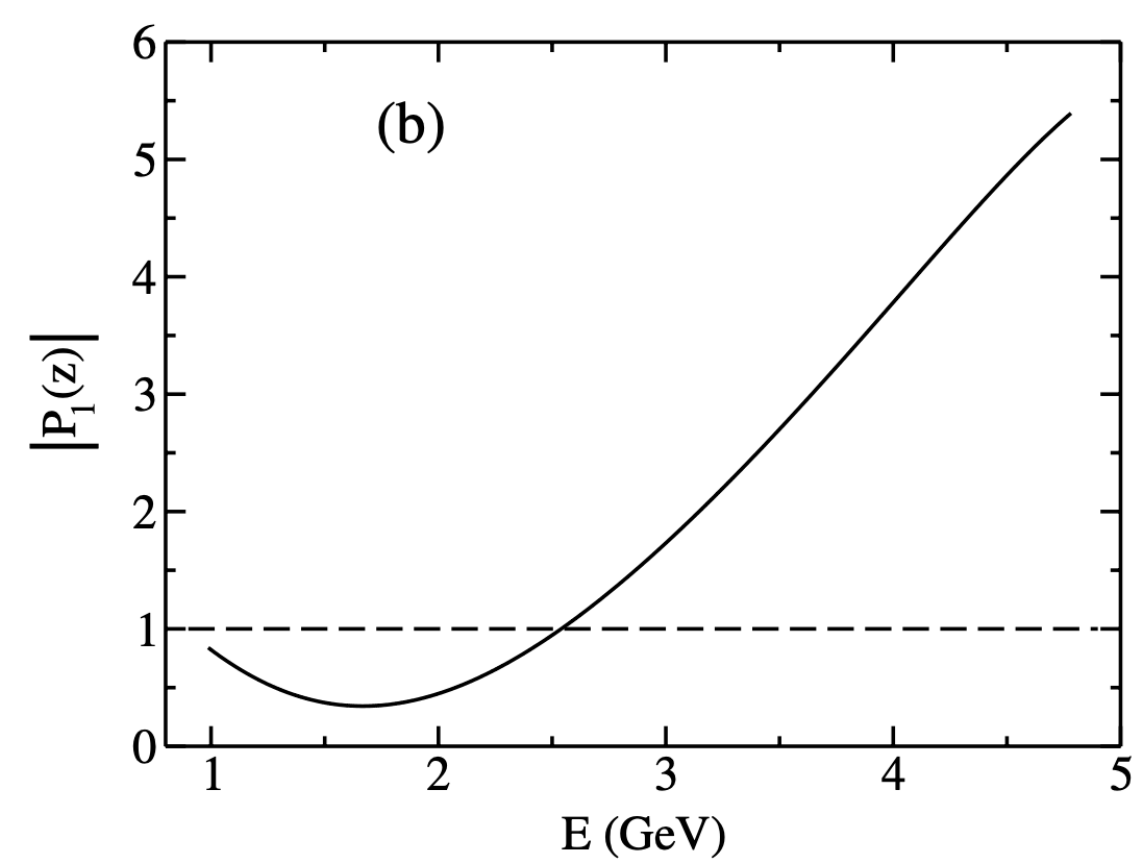}
\caption{ (a) Comparison between $\left \vert \bar{A}_1(s_0)\right \vert$ (Eq.~(\ref{A6S0b}), solid blue line) and  $\left \vert \bar{A}_1^F \right \vert \equiv  \left \vert P_1(z) \bar{A}_1(s_0) \right \vert$ (dashed red line) with~$z~=~E-~2~m_K$ and $E=\sqrt{s_0}$; (b)  plot of $\left \vert P_1(z) \right \vert$ (Eq.~(\ref{P1}), solid line).
}\label{F7ModA1}
\end{center} \end{figure}

The $R_j$ results shown in Table~\ref{rj} tell us that the contribution to the amplitude with two kaons in  isoscalar $S$ wave is very important.
It is instructive to plot the modulus of the modified $\bar A_1(s_0)$ contribution, $\left \vert  \bar A_1^F \right \vert \equiv  \left \vert P_1(z) \bar A_1(s_0) \right \vert $ and to compare it to the modulus of $\bar A_1(s_0)$, this is done in Fig.~\ref{F7ModA1}(a).
The fit to the data requires a reduction of  $|\bar A_1|$  below 2.5~GeV and to an increase above  which is  done by  $|P_1(z)|$  as seen in Fig.~\ref{F7ModA1}~(b).
Our strange kaon form factor is too large in the energy range below 2.5~GeV and too small above.
 Besides this strange form factor $\Gamma_2^{s*} (s_0)$ the $\bar B^0$  to $K^0$ transition form factor $F_0^{\bar B^0 \bar K^0}(s_0)$ enters the expression of $\bar A_1$ [Eq.~(\ref{A6S0b})]
 and the product of these two form factors is constrained by the data.
 The  $F_0^{\bar B^0 \bar K^0}(s_0)$ given by Eq.~(\ref{F0B0K0})  and evaluated in Ref.~\cite{Ball_PRD71_014015} from light cone sum rules is in good agreement with that recently calculated in a fully relativistic lattice QCD approach~\cite{PRD_107.014510}.
 This can be seen comparing the values given by the parametrization~(\ref{F0B0K0}) to those of the curve of Fig.~16 and Table~VI of Ref.~\cite{PRD_107.014510}.

\begin{figure}[h]  \begin{center}
\includegraphics[scale=0.265]{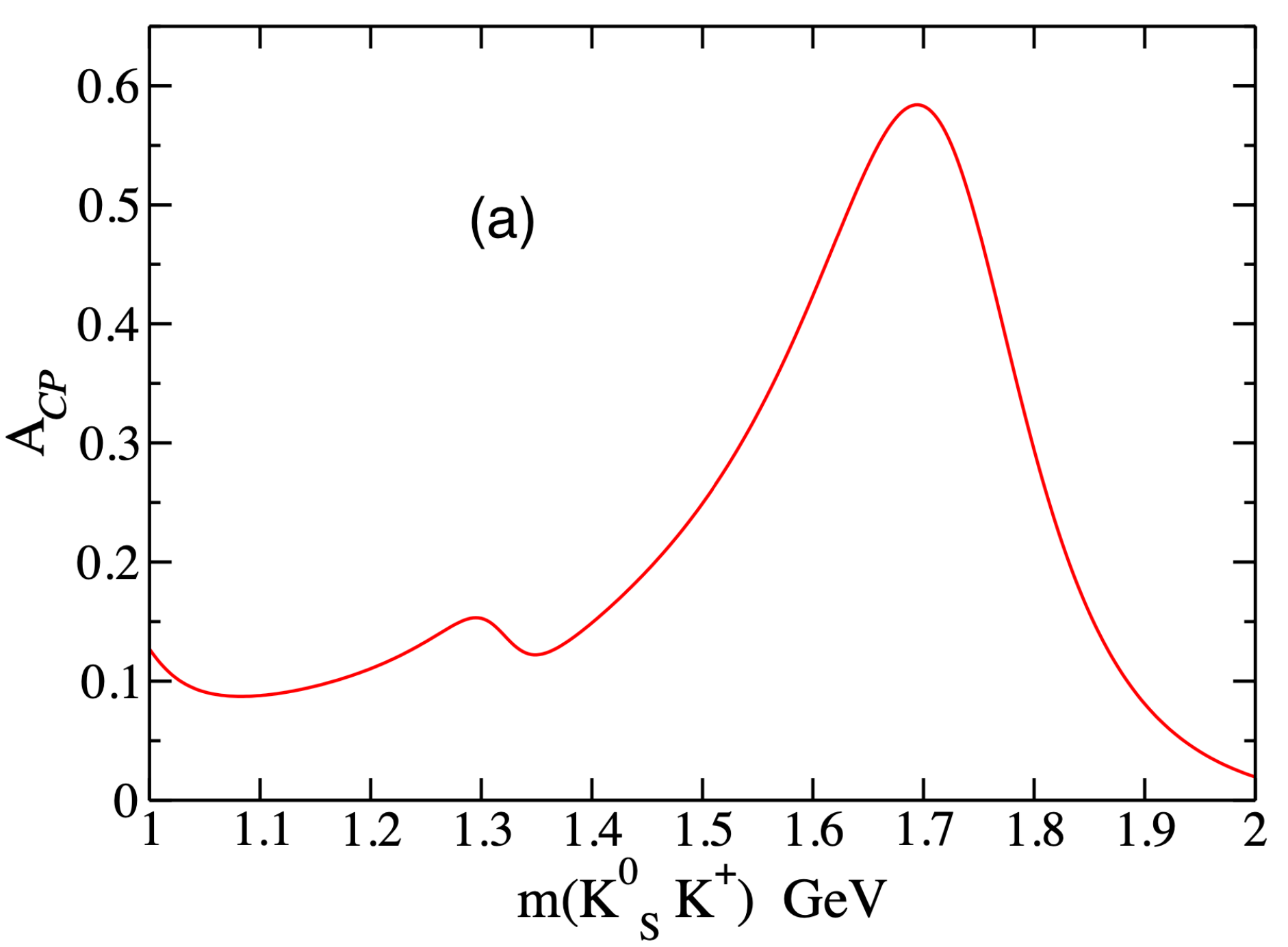}
\includegraphics[scale=0.265]{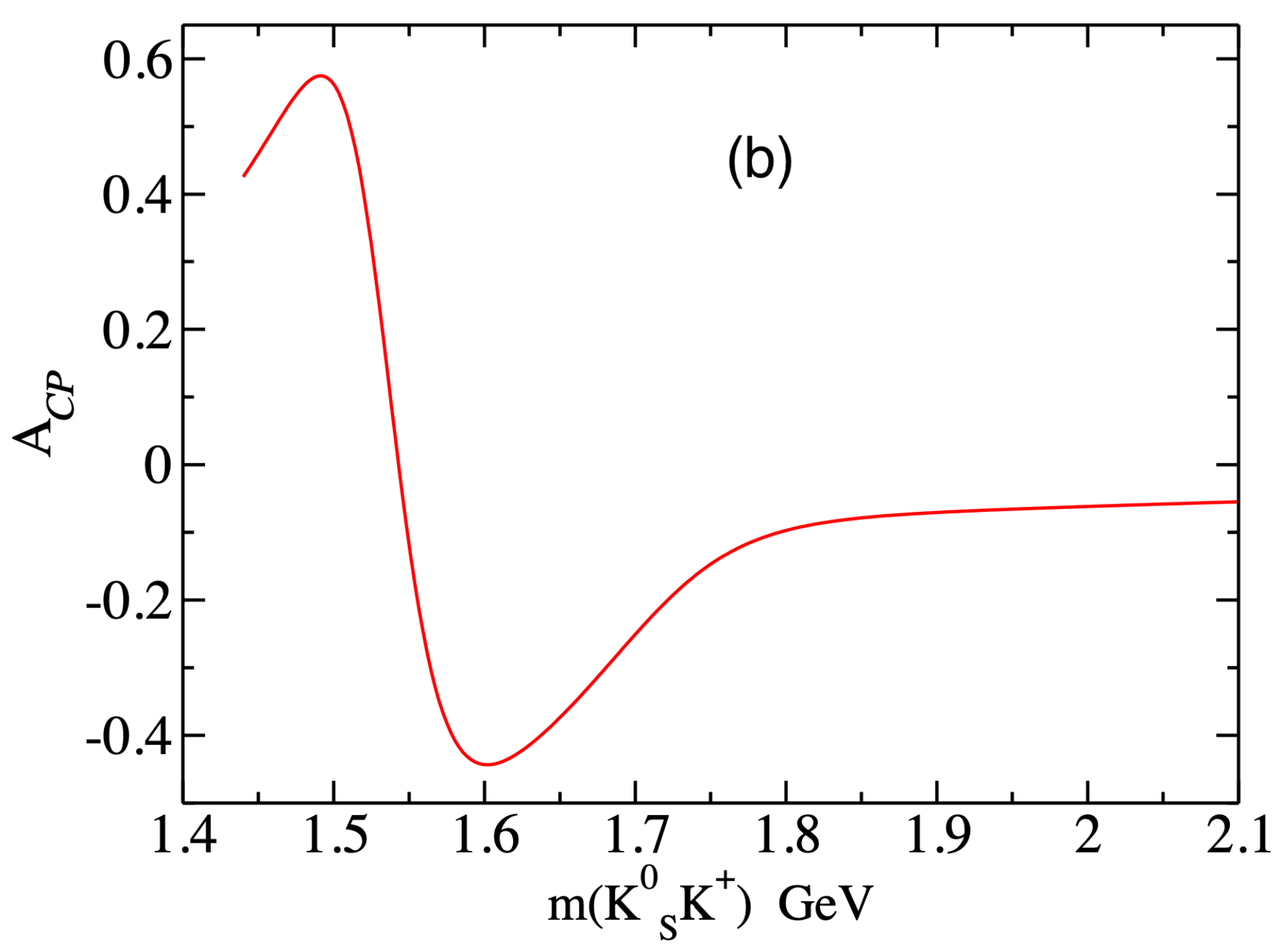}
\caption{Asymmetry $A_{{CP}}(s_0,s_+)$ [Eq.~(\ref{ACPdef})] as a function of $m_+$ for: (a)  $m_0=$~3.5~GeV, (b)  $m_0 = $~2~GeV.
In plot~(a), the $a_2(1320)^+$,  present in the modified $A_5$ contribution, could be responsible for the bump around~1.3~GeV.
 The maximum at $\sim1.7$~GeV can be related to an interplay between the modified $A_1$ and $A_2$ terms where the two-kaon states are in $S$ wave.
}\label{ACPfig}
\end{center} \end{figure}
The Dalitz-plot dependence of {\it CP} asymmetry in the framework of a QCDF model for the $B^\pm~\to~K^\pm~K^+ K^-$ has been compared to LHCb~\cite{PRL111_101801} and \textit{BABAR}~\cite{PRD85_112010} data in Ref.~\cite{KKK}.
In a recent publication~\cite{PRD108_012008} the LHCb Collaboration has reported measurement of  {\it CP} asymmetries in charmless three-body decays of $B^\pm$.
They have shown their distributions as a function of the three-body phase space
and have interpreted them as possibly arising from rescattering and resonance interference effects.
 
For  the $B^0 \to K^0_S K^+ K^- $ decays the {\it CP} asymmetry in the Dalitz plot can be defined using Eqs.~(\ref{d2Brbar}) and~(\ref{d2Br}) as follows:
\be
\label{ACPdef}
A_{{CP}}(s_0,s_+)=\frac{\frac{d^2{\rm Br(\bar{ B^0})}}{ds_+ds_0}-
\frac{d^2{\rm Br(B^0)}}{ds_-ds_0}}{\frac{d^2{\rm Br(\bar{ B^0})}}{ds_+ds_0}+
\frac{d^2{\rm Br(B^0)}}{ds_-ds_0}}.
\ee
In a large part of the Dalitz plot  the $A_{CP}(s_0,s_+)$ values are rather small but there are some regions with $m_0~\gtrsim~1.7$~GeV where they can be sizable.
For instance, in Fig.~\ref{ACPfig}~(a) we show a plot of  $A_{CP}(s_0,s_+)$ as a function of $m_+$ for $m_0$=3.5 GeV. 
Here we encounter large positive  asymmetry values.
A small maximum at $m_+~\simeq~1.3$~GeV can be attributed to an influence of the resonance $a_2(1320)$ present in the two-kaon-D-wave contribution of the modified $A_5$ term.
The second higher maximum has no direct resonant character.
It can be related to an interplay of the two-kaon-S-wave contributions of the modified $A_1$ and $A_2$ terms.
For other values of $m_0$ and $m_+$, for example, as seen in Fig.~9~(b), at $m_0= 2$~GeV and for 1.55~GeV$\lesssim m_+\lesssim~2.15 $~GeV a significant negative asymmetry is found while for 1.4~GeV$<m_+<~1.55$~GeV the asymmetry is large and positive.
Also for $m_0$=4~GeV and in the whole range of the kinematically allowed $m_+$ values from 1~GeV to 3.4~GeV the $CP$ asymmetry is large and positive for $m_+$ below 1.55~GeV and negative above.
After integrations on $s_0, s_+$ in the Dalitz plot and as noted below Eq.~(\ref{ACP}),  the total CP asymmetry, equal to  $-$0.11\%, is small.

The time dependent asymmetry $A_{{CP}}(\Delta t)$ is usually written as 
\be \label{ACPt}
A_{{ CP}}(\Delta t)=-~C \cos(\Delta m_d \Delta t)+S \sin(\Delta m_d \Delta t),
\ee
where $\Delta t$ is defined as the time interval between the decays of the $B^0$ mesons coming from the $\Upsilon(4S)$ state, while $C$ and $S$ are coefficients
which can depend on the Dalitz plot variables like $s_0$ and $s_+$. 
These coefficients can be calculated as ratios of integrals over some parts of the Dalitz plot, namely\footnote{See Eqs.~(\ref{BrB0time}) to~(\ref{S}).} $C=-I_C/D$ and $S=I_S/D$,
where
\be 
\label{IC}
I_C=\int d s_0\, ds_+ (\,|\bar{A}(s_0,s_-,s_+)|^2 - |A(s_0,s_+,s_-)|^2\,),
\ee 
\be 
\label{ID}
D=\int d s_0\, ds_+ (\,|\bar{A}(s_0,s_-,s_+)|^2 + |A(s_0,s_+,s_-)|^2\,),
\ee
and
\be 
\label{IS}
I_S=2 \int  d s_0\, ds_+ \,{\rm Im}\,[\,{\rm e}^{-2 i \beta} \bar{A}(s_0,s_-,s_+)\, A^*(s_0,s_+,s_-)\,].
\ee
In Eq.~(\ref{IS}) the angle $\beta$ is that of the  unitarity triangle~\cite{PDG2022}.
One can see that, when the $s_0, s_+$ integration is performed over the  full Dalitz plot, the coefficient $C$ is equal to the {\it CP} asymmetry with a minus sign, $C=-A_{{CP}}$.

Using $\sin(2 \beta)=0.699$~\cite{PDG2022} and integrating over $s_+$ and for  three specific ranges of $s_0$, we obtain the $C$ and $S$ values given in  Table~\ref{CS}.
One notices a sign flip of the coefficient $S$ when going from the $s_0$ range dominated by the $\phi(1020)$ meson contribution to the $s_0$ range outside of $\phi(1020)$.
The change of the $S$ sign is related to the presence of an additional minus sign in the amplitude $A_3$ with respect to the corresponding $\bar{A_3}$ amplitude. 
The charge symmetry of the $P$-wave $K^+K^-$ amplitudes is responsible for that effect. 
The numerical values of the time dependent  {\it CP}-asymmetry parameters are in qualitative agreement with the experimental results of the \textit{BABAR} Collaboration presented in Fig.~18 of Ref.~\cite{PRD85_112010}.
The $S$ value in the $\phi$ region, 0.53, is compatible with that of \textit{BABAR}, 0.66~$\pm$~0.17~$\pm$~0.07, given in Table~XIII of Ref~\cite{PRD85_112010}.

\begin{table}
\caption{{\it CP}-asymmetry parameters $C$ and $S$ defined in Eq.~(\ref{ACPt}) and calculated from Eqs.~(\ref{IC}), (\ref{ID}) and (\ref{IS}) integrated over the full $s_+$ range and over the specific $s_0$ range.}
\begin{center}
\begin{tabular}{ccc}
\hline
\hline
$\sqrt{s_0}$ range (GeV) &~~  $C=-A_{{CP}}$~ (\%) &~~ $S$ \vspace*{0.2cm} \\ 
\hline
$1.01< \sqrt{s_0}< 1.03$ & $-1.13$ & $+0.53$\vspace*{0.2cm} \\   
$\sqrt{s_0}< 1.01$ and $\sqrt{s_0}> 1.03$ & $0.43$ & $-0.61$\vspace*{0.2cm} \\ 
full $ s_0$ range & $0.17$ &$-0.42$ \vspace*{0.2cm} \\  
\hline
\hline
\end{tabular}
\end{center}
\label{CS}
\end{table}

\section{Summary and concluding remarks} \label{conclusions}

In view of further amplitude analyses, in particular by LHCb and Belle~II  Collaborations,
we have derived a $ B^0 \to K^0_S K^+ K^-$ decay amplitude in a quasi-two-body QCDF framework.
Our derivation follows that developed for the study of {\it CP} violation in the $B^{\pm}\to \pi^+ \pi^- \pi^{\pm}$ decays~\cite{DedonderPol}.
The dominant parts of the decay amplitude are calculated in terms of kaon form factors or $B^0$ to two kaons transition functions which describe the final state two-body resonances and their interferences.
Unitarity constraints are satisfied when  two of the three kaons are in a scalar state.
The kaon form factors and transition functions entering this amplitude are similar to those introduced in the Dalitz plot studies of the $D^0 \to K_S^0 K^+ K^-$ decays in a factorization approach~\cite{JPD_PRD103}, the final kaon states being identical.
However, here, the larger phase space tests our model over a wider energy range.
The kaon-kaon interactions in the $S$, $P$, and $D$ waves  are taken into account.

Starting from the effective weak decay Hamiltonian~\cite{Ali1998,Beneke:2001ev}, a QCDF derivation of the full amplitude within a quasi-two-body framework can be performed. 
The different terms [see Eqs.~(\ref{TpK0KK})] appear as products of short distance contributions, sums which depend on effective Wilson coefficients [see Eq.~(\ref{ajp})],  times long distance ones given by kaon form factors or parametrized with $\bar B^0 $ to $\bar K^0 K^+$ transition functions.
Some parts of the amplitude, where the formation of the final  $K^+ K^-$  takes place via an implicit or explicit $d \bar d$ quark pair, are expected to lead to small contributions.
We have neglected these OZI~\cite{OZI} suppressed terms.

The dominant part of the full amplitude has five components and 
our model reproduces  well the Belle~\cite{PRD82_073011} and \textit{BABAR}~\cite{PRD85_112010} Collaborations data.
With 13 strong interaction free parameters modifying the five terms of our amplitude, we fit the 422 observables consisting of the total branching fraction together with the Dalitz-plot  projections of Belle and \textit{BABAR} with a $\chi^2$ of 583.6 which leads to a $\chi^2/ndf$ of 1.43.

The largest contribution to the branching fraction, 83.0 \% of the total as seen in Table~\ref{rj},  comes from the modified $\bar A_1(s_0,s_-, s_+)$ and  $A_1(s_0,s_+, s_-)$ terms where the $K^+K^-$ pairs are in a scalar-isoscalar state (see the penguin diagrams of Fig.~\ref{F2}~(b)).
These terms are proportional to the strange  scalar-isoscalar form factor $\Gamma_2^s(s_0)$ receiving a large contribution from the $f_0(980)$, $f_0(1370)$ and $f_0(1500)$ resonances (see Fig.~\ref{ModGam2s}).
The dominance of the $f_0$-resonance contributions was also found in the data analyses of the  Belle~\cite{PRD82_073011} and \textit{BABAR}~\cite{PRD85_112010} Collaborations.

The best fit is obtained if the $\bar A_1(s_0,s_-, s_+)$ and  $A_1(s_0,s_+, s_-)$  terms [Eqs.~(\ref{A6S0b}), (\ref{A6S0})] are multiplied by the phenomenological complex polynomial $P_1(z)$ with $z=\sqrt{s_0}- 2 m_K$ (see Eq.~(\ref{P1}), Fig.~\ref{F7ModA1}~(b) and  Table~\ref{parameters}).
It leads to a reduction of  $|\bar A_1|$ and $|A_1|$  below 2.5~GeV and to an increase above [see Fig.~\ref{F7ModA1}~(a)].
Within our approach, and for a given $\bar B^0$  to $K^0$ transition form factor $F_0^{\bar B^0 \bar K^0}(s_0)$,  the fit to the ${B^0}\to K^0_S K^+ K^-$ Belle~\cite{PRD82_073011} and {\it BABAR}~\cite{PRD85_112010} data
would require for the modulus of our strange kaon form factor $\Gamma_2^s(s_0)$,  a smaller (larger) value for $\sqrt{s_0}$ below (above) 2.5~GeV.

The next important mode, with a branching fraction equal to 18.3\% of the total is mainly the $\phi$~$K^0_S$ one plus some small $\omega$~$K^0_S$ and $\rho$~$K^0_S$ arising from the modified $\bar A_3(s_0, s_-, s_+)$ and $ A_3(s_0, s_+, s_-)$ amplitudes.
The dominant part with the $\phi(1020)$ contribution comes from the term proportional to $w_s$~[Eq.~(\ref{ws})].
The  parameters, for the $P$-wave form factors $F_s^{K^+K^-}(s_0)$ and $F_u^{K^+K^-}(s_0)$ [Eqs.~(\ref{FuK+K-I}) and~(\ref{FsK+K-0})] have been determined in Ref.~\cite{Bruch_2005} using vector dominance, quark model assumptions and isospin symmetry.
The best fit  requires these $\bar A_3$ and $ A_3$ [Eqs.~(\ref{M35P0b}) and~(\ref{M35P0})] contributions to be renormalized by a real parameter $P_3$ which is, however, close to 1 (see Table~\ref{parameters}).

The modified terms $\bar{A}_2(s_{0},s_{-},s_{+})$ and ${A}_2(s_{0},s_{+},s_{-})$ with two kaons in a $S$ wave of isospin 1, have a branching faction of 6.2~\% of the total.
Their long distance part depends upon the function $G_1(s_\pm)$ whose calculation, given by Eqs.~(104) to~(111) of Ref.~\cite{JPD_PRD103}, is based on the  $\pi \eta$- and $K \bar K$-channel model of the  $a_0(980)$ and $a_0(1450)$ resonances built in~Refs.~\cite{AFLL, AFLL2}.
To obtain a good fit, we found necessary to adjust for the $G_1(s)$ function the $r_2$  coupling to the $\bar K^0 K^+$ state and to multiply the $\bar{A}_2$ and ${A}_2$ terms, [Eqs.~(\ref{A2b}) and~(\ref{A2})] by the phase factor ${\rm e}^{i\phi_2}$ (see Table~\ref{parameters}).
The contributions of the $a_0$ resonances were not introduced  in the  Belle~\cite{PRD82_073011}  and \textit{BABAR}~\cite{PRD85_112010} Collaboration analyses.

The  remaining amplitudes $\bar{A}_4 (s_{0},s_{-},s_{+})$ and ${A}_4 (s_{0},s_{+},s_{-})$ 
 (contributions of the $\rho(770)$, $\rho(1450)$, and $\rho(1700)$ resonances and renormalized by the real parameter $P_4$),  $\bar{A}_5 (s_{0},s_{-},s_{+})$ and ${A}_5 (s_{0},s_{-},s_{+})$ ($D$-wave saturated by the $a_2(1320)^+$ resonance and multiplied by the real parameter $P_5$) give small branching fractions of the order of 0.1\%.

For $K^+ K^-$ effective masses  above 1.7~GeV our model predicts large {\it CP} asymmetries in the Dalitz plot, as can be seen in Figs.~\ref{ACPfig}~(a) and~(b).
We have also calculated the values of the time dependent  {\it CP}-asymmetry parameters and
in the $\phi(1020)$ region, the value of the $S$~parameter, 0.53, agrees, within errors, with that obtained by \textit{BABAR} analysis~\cite{PRD85_112010}.

The charmless three-body  $ B^0 \to K^0_S K^+ K^-$ decay data provides 
information on the weak  interactions and can also be useful for  better knowledge of the kaon-kaon strong interactions.
Based on our model one can build a parametrization that can be implemented in experimental Dalitz plot analyses.
 Dalitz-plot amplitude analysis of several charmless three-body $B$-meson decays can lead to a better understanding on the origin of {\it CP} asymmetry.

\section*{Acknowledgements}
\label{ACKN}
{We are deeply indebted to Bachir Moussallam for very profitable correspondence and for the communication of the results of his calculation of scalar form factors.
We thank Agnieszka Furman for her participation in the early stage of this work.
We also acknowledge helpful discussions with Emi Kou, Mathews Charles and Eli Ben-Haim. 
This work has been partially supported by a grant from the French- Polish exchange program COPIN/CNRS-IN2P3, collaboration 23-155.

\appendix

\section{\mbox{\boldmath $B^0$-$\bar {B}^0$} mixing and time-dependent decay rate}
\label{B0B0bmixingtimedep}

The quantum mechanical formalism for neutral particle-antiparticle oscillations and {\it CP} violation  has been studied and presented in the book  of I.~I.~Bigi and A.~I.~Sanda~\cite{CP_Violation2000} (herafter cited as BS).
Recent developments on the  {$B^0$-$\bar {B}^0$ mixing can  be found in the review by O.~Schneider~\cite{B0_B0barmixing} in the 2022 Review of Particle Physics~\cite{PDG2022}.
In this appendix, following BS we show how one can derive Eq.~(3) of  the \textit{BABAR} study of {\it CP} violation in Dalitz-plot analyses of the charmless hadronic $B^0 \to K^0_S K^+ K^-  $ decay~\cite{PRD85_112010}.
We also give the derivation of Eqs.~({\ref{d2Brbar}),~(\ref{d2Br}) and~(\ref{ACPt}).

\subsection{\mbox{\boldmath $B^0$-$\bar {B}^0$} mixing}
\label{B0B0bmixing}
The expressions for the time evolution of  $B^0$ and $\bar B^0$  states are given by (see Eqs.~(6.47) and (6.48) of BS)
 \bqa  \label{TimeEvolution}
|B^0(t)\rangle&=& f_+(t)||B^0(t)\rangle+\frac{q}{p} f_-(t)||\bar{B}^0(t)\rangle, \nonumber \\
|\bar{B}^0(t)\rangle&=& f_+(t)||\bar{B}^0(t)\rangle+\frac{p}{q} f_-(t)||{B}^0(t)\rangle
\eqa
with\footnote{$\Delta M_{B^0}$ is denoted as $\Delta m_d$ and $t$ as $\Delta t$ in Sec.~\ref{results}.}  
\be\label{fpm}
f_{\pm}(t)=\frac{1}{2} {\rm e}^{-iM_{S}t} {\rm e}^{-\frac{1}{2}\Gamma_{S}t}\left( 1 \pm {\rm e}^{-i\Delta M_{B^0 }t} {\rm e}^{-\frac{1}{2}\Delta \Gamma_{B^0 }t}\right).
\ee
In Eq.~(\ref{fpm}) $\Delta M_{B^0 }\equiv M_L-M_S$ and $\Delta \Gamma_{B^0}\equiv \Gamma_S-\Gamma_L  \ll  \Gamma_S+\Gamma_L$, i. e.  $\Gamma_S \simeq \Gamma_L$ (see Eq.~(11.2) of BS).
Here $M_L, \Gamma_L$ and $M_S, \Gamma_S $ correspond to the masses and widths of the long-life and short-life $B^0$ states, respectively.
The time-dependent differential decay rate can be written  as 
\be \label{d2BrBS}
\frac{d\Gamma}{ds_+ds_0dt}=\frac{{\rm e}^{-\Gamma_{B^0 }t}} {32(2\pi)^3m_{B^0}^3} \frac{G(t)}{4\tau_{B^0}},
\ee
where $\tau_{B^0}$ equal to  $1/\Gamma_{B^0}$ is the neutral B meson lifetime.
Applying the BS master equations (11.15) to (11.22) one obtains for the  $ B^0 \to K^0_S K^+ K^- $ decay (Eq.~(11.58) of BS):
\be \label{GfB03K}
G(t)= |A|^2 \left [ 1+|\bar {\rho}|^2 + \left (1-|\bar {\rho}|^2\right)  \cos({\Delta M_{B^0 }t}) - 2 {\rm Im} \left(\frac{q}{p} \bar{A} {A}^* \right)\sin({\Delta M_{B^0}t})\right ],
\ee
where  $A=\sum_{i=1}^5 A_i$ is the  $ B^0 \to K^0_S K^+ K^- $ decay amplitude and (see Eq.~(6.49) of BS)
\be \label{barrho}
\bar {\rho}=\frac{\bar{A}}{A}=\frac{1}{\rho},
\ee
$\bar {A}$ being the  $ \bar{B}^0 \to K^0_S K^- K^+ $ decay amplitude.
For the definition of $p$ and $q$ see e.g. Eqs.~(6.22) to (6.25) of BS.
In the $B^0$ case one has (Eq. (11.45) of BS and Ref.~\cite{PDG2022}),
\be \label{poverqB0}
\frac{q}{p} = \frac{V_{tb}^* V_{td}}{V_{tb}V_{td}^*}\simeq {\rm e}^{-2i \beta},
\ee
where $\beta$ is one of the angles of the CKM triangle.
From Eqs.~(\ref{GfB03K}) and (\ref{barrho}) one gets
\be \label{TimeDepB0}
G(t)= |A|^2 + |\bar{A}|^2  +\left (\ |A|^2 - |\bar{A}|^2 \right) \cos({\Delta M_{B^0 }t}) - 2 {\rm Im} \left ({\rm e}^{-2i \beta} \bar{A} A^*\right) \sin({\Delta M_{B^0}t}).
\ee

Following Eqs.~(\ref{d2BrBS}) and (\ref{TimeDepB0}) the time dependent double differential branching fraction of the $B^0$ decay, with $Br=\Gamma / \Gamma_{B^0}$ and $N_{Br}\equiv [32(2\pi)^3m_{B^0}^3 \Gamma_{B^0})]^{-1}$, reads  (with the replacement of $t$ by $\Delta t$)
\bqa \label{d2BrB0time}
\hspace*{-0.3cm}\frac{d^2 Br(B^0)}{ds_+\ ds_0\ d\Delta t}&=&N_{Br}\frac{{\rm e}^{-\Gamma_{B^0}|\Delta t|}}{4 \tau_{B^0}} \nonumber \\
&\times&\ \hspace*{-0.3cm}\left [|A|^2 + |\bar{A}|^2  +\left (|A|^2 - |\bar{A}|^2 \right) \cos({\Delta M_{B^0 }\Delta t}) - 2 {\rm Im} \left({\rm e}^{-2i\beta}\bar{A} A^*\right) \sin({\Delta M_{B^0} \Delta t})\right]\ 
\eqa
and that of the $\bar{B}^0$
\bqa \label{d2BrB0bartime}
\hspace*{-0.3cm}\frac{d^2 Br(B^0)}{ds_+\ ds_0\ d\Delta t}&=&N_{Br}\frac{{\rm e}^{-\Gamma_{B^0}|\Delta t|}}{4 \tau_{B^0}} \nonumber \\
&\times&\ \hspace*{-0.3cm}\left [|\bar A|^2 + |{A}|^2  +\left (|\bar A|^2 - |{A}|^2 \right) \cos({\Delta M_{B^0 }\Delta t}) + 2 {\rm Im} \left( {\rm e}^{-2i\beta}\bar{A} A^*\right) \sin({\Delta M_{B^0} \Delta t})\right]\ 
\eqa
This shows the agreement of Eqs.~(\ref{d2BrB0time}) and~\ref{d2BrB0bartime}) with Eq.~(3) of  Ref.~\cite{PRD85_112010} for $w=0$.
Integrating over the time from minus to plus infinity and with,
\be \label{intcos}
\int^{+\infty}_{-\infty} d\Delta t \ \frac{{\rm e}^{-\Gamma_{B^0}|\Delta t|}}{4 \tau_{B^0}}  \cos({\Delta M_{B^0 }\Delta t})=\frac{1}{2}\frac{1}{\left(\frac{\Delta M_{B^0}}{\Gamma_{B^0}}\right)^2+1},
\ee
and
\be \label{intsin}
\int^{+\infty}_{-\infty} d\Delta t \ \frac{{\rm e}^{-\Gamma_{B^0}|\Delta t|}}{4 \tau_{B^0}}  \sin({\Delta M_{B^0 }\Delta t})=0,
\ee
one obtains from Eqs.~(\ref{d2BrB0time}) and ~(\ref{d2BrB0bartime})
\be \label{d2BrA}
\frac{d^2{\rm Br(B^0)}}{ds_+ds_0}=N_{Br}
[\,(1-x)|A(s_0,s_+,s_-)|^2+x |\bar{A}(s_0,s_+,s_-)|^2\,],
\ee
and
\be \label{d2BrbarA}
\frac{d^2{\rm Br(\bar{ B^0})}}{ds_+ds_0}=N_{Br}
[(1-x)|\bar{A}(s_0,s_-,s_+)|^2+x |A(s_0,s_-,s_+)|^2],
\ee
where (introducing here the $w$ dependence) $x= \frac{1}{2}~\Big[~1-\frac{1-2 w}{(\Delta M_{B^0 } /\Gamma_{B^0})^2+1}~\Big]$.
Eqs.~(\ref{d2BrA}) and ~(\ref{d2BrbarA}) correspond to   Eqs.~(\ref{d2Br}) and ~(\ref{d2Brbar}).

\subsection{Time dependent asymmetry \mbox{\boldmath $A_{{CP}}( t)$}}
\label{ACPtsec}

Integrating over $s_+$ and $s_0$ and denoting by $\tilde {B}_r$ the total branching fraction without $B^0$-$\bar {B}^0$ mixing, one obtains from Eqs.~(\ref{d2BrB0time}) and (\ref{d2BrB0bartime})  for the $B^0$ decay 
\bqa \label{BrB0time}
Br_{B^0}(\Delta t)&=&\frac{{\rm e}^{-\Gamma_{B^0 }|\Delta t|}}{2 \tau_{B^0}}
\Big [\tilde{Br}_{B^0} + \tilde{Br}_{\bar{B}^0} +\left (\tilde{Br}_{B^0}-\tilde{Br}_{\bar{B}^0} \right) \cos({\Delta M_{B^0 }|\Delta t|})   \nonumber \\  &-& 2 N_{Br}\int \int {\rm Im} \left( {\rm e}^{-2i\beta}\bar{A} A^*\right)\sin({\Delta M_{B^0} |\Delta t|})\ ds_+ds_0\Big ],
\eqa
and for the $\bar{B}^0$ decay
\bqa \label{BrbarB0time}
Br_{\bar{B}^0}(\Delta t)&=&\frac{{\rm e}^{-\Gamma_{B^0}|\Delta t|}}{2 \tau_{B^0}}
\Big [\tilde{Br}_{\bar{B}^0} + \tilde{Br}_{{B}^0} +\left (\tilde{Br}_{\bar{B}^0}-\tilde{Br}_{{B}^0} \right) \cos({\Delta M_{B^0 }\Delta t})   \nonumber \\  &+& 2 N_{Br}\int \int {\rm Im} \left( {\rm e}^{-2i\beta}{\bar A} {A}^* \right)\sin({\Delta M_{B^0} \Delta t})\ ds_+ds_0\Big ].
\eqa
The time dependent asymmetry $A_{{CP}}( \Delta  t)$ defined as
\be
A_{{CP}}( \Delta t)=\frac{Br_{\bar{B}^0}( \Delta t)-Br_{B^0}( \Delta t)} {Br_{\bar{B}^0}( \Delta t)+Br_{B^0}( \Delta t)}
\ee
 is usually written as 
\be \label{ACPtA}
A_{{CP}}( \Delta t)=-\ C \cos(\Delta M_{B^0}  \Delta  t)+S \sin(\Delta M_{B^0} \Delta  t).
\ee
From Eqs.~(\ref{BrB0time}), (\ref{BrbarB0time}) 
one obtains\footnote{As seen from  Eqs.~(\ref{d2BrA}) and ~(\ref{d2BrbarA}) the $B^0$-$\bar {B}^0$ mixing cancels when adding ${B}^0$ and $\bar B^0$ branching fractions with mixing.}
\be \label{C}
C=-\frac{{Br}_{\bar{B}^0} - {Br}_{B^0}}{Br_{\bar{B}^0}+Br_{B^0}} = - A_{{ CP}},
\ee
and
\be \label{S}
S= \frac{2\int \int {\rm Im} \left ( {\rm e}^{-2i\beta}\bar {A} A^*\right ) ds_+ds_0}{ Br_{\bar{B}^0}+Br_{B^0}}.
\ee
Equations~(\ref{BrB0time}) to~(\ref{S}) are equivalent to Eqs.~(\ref{ACPt}) to~(\ref{IS}).

\end{document}